\documentclass[twocolumn]{aastex631}
\shortauthors{Yu et al.}
\usepackage{amsmath}
\usepackage{CJK}
\usepackage[T1]{fontenc}
\usepackage[utf8]{inputenc}

\newcommand{\lthreezero}{$L_{\rm 3000}$}
\newcommand{\lledd}{$L/L_{\rm Edd}$}

\newcommand{\lbol}{$L_{\rm bol}$}

\newcommand{\msun}{$M_{\odot}$}

\newcommand{\sigmadho}{$\sigma_{\mathrm{DHO}}$}

\newcommand{\sigmanoise}{$\sigma_{\mathrm{\epsilon}}$}
\newcommand{\tauperturb}{$\tau_{\mathrm{perturb}}$}
\newcommand{\taudecay}{$\tau_{\mathrm{decay}}$}

\newcommand{\taurise}{$\tau_{\mathrm{rise}}$}

\newcommand{\lambdarf}{$\lambda_{\mathrm{RF}}$}

\newcommand{\mbh}{$M_{\rm {BH}}$}

\renewcommand{\arraystretch}{1.1} 
\newcommand{\fhmgii}{$\rm FWHM_{Mg\,\textsc{ii}}$}
\newcommand{\rmAA}{$\rm \AA$}

\renewcommand{\edit}[1]{{#1}}
\defcitealias{yu2022b}{Y22}
\defcitealias{moreno2019}{M19}
\defcitealias{kasliwal2017}{K17}
\begin{document}
\begin{CJK*}{UTF8}{gbsn}

\title{Examining AGN UV/Optical Variability Beyond the Simple Damped Random Walk. II. \\Insights from 22 Years Observations of SDSS, PS1 and ZTF}

\correspondingauthor{Weixiang Yu}
\email{wyu@ubishops.ca}

\author[0000-0003-1262-2897]{Weixiang Yu~(于伟翔)}
\affiliation{Department of Physics, Drexel University, 32 S.\ 32nd Street, Philadelphia, PA 19104, USA}
\affil{Department of Physics \& Astronomy, Bishop's University, 2600 rue College, Sherbrooke, QC, J1M 1Z7, Canada}

\author[0000-0002-1061-1804]{Gordon T. Richards}
\affiliation{Department of Physics, Drexel University, 32 S.\ 32nd Street, Philadelphia, PA 19104, USA}

\author[0000-0001-8665-5523]{John~J.~Ruan}
\affil{Department of Physics \& Astronomy, Bishop's University, 2600 rue College, Sherbrooke, QC, J1M 1Z7, Canada}

\author[0000-0001-7416-9800]{Michael S. Vogeley}
\affiliation{Department of Physics, Drexel University, 32 S.\ 32nd Street, Philadelphia, PA 19104, USA}

\author[0000-0002-8686-8737]{Franz~E.~Bauer}
\affil{Instituto de Alta Investigaci{\'{o}}n, Universidad de Tarapac{\'{a}}, Casilla 7D, Arica, Chile}

\author[0000-0002-3168-0139]{Matthew J. Graham}
\affiliation{Department of Physics, Math, and Astronomy, California Institute of Technology, Pasadena, CA, 91125, USA}


\begin{abstract}
A damped random walk (DRW) process is often used to describe the temporal UV/optical continuum variability of active galactic nuclei (AGN). However, recent investigations have shown that this model fails to capture the full spectrum of AGN variability.
In this work, we model the 22-year-long light curves of $21,767$ quasars, \edit{spanning the redshift range $0.28 < z <  2.71$,} as a noise-driven damped harmonic oscillator (DHO) process. The light curves, \edit{in the optical $g$ and $r$ bands,} are collected and combined from \edit{the Sloan Digital Sky Survey}, \edit{the Panoramic Survey Telescope and Rapid Response System}, and \edit{the Zwicky Transient Facility}. A DHO process can be defined using four parameters, two for describing its long-term behavior/variability, and the other two for describing its short-term behavior/variability. We find that the best-fit DHO model describes the observed variability of our quasar light curves better than the best-fit DRW model. Furthermore, the best-fit DHO parameters exhibit correlations with the rest-frame wavelength, the Eddington ratio, and the black hole mass of our quasars. 
Based on the power spectral density shape of the best-fit DHOs and these correlations, we suggest that the observed long-term variability of our quasars can be best explained by accretion rate or thermal fluctuations originating from the accretion disk, and the observed short-term variability can be best explained by reprocessing of X-ray variability originating from the corona. The additional information revealed by DHO modeling emphasizes the need to go beyond DRW when analyzing AGN light curves delivered by next-generation wide-field time-domain surveys.
\end{abstract}


\section{Motivation}\label{sec:motivation}
The UV/optical continuum luminosity of active galactic nuclei (AGN)\footnote{In this manuscript, we will use AGN and quasar interchangeably without making a distinction based on, e.g., luminosity or radio emission.} typically exhibit stochastic variability on time scales ranging from days to decades~\citep{vandenberk2004, sesar2007}.
It is now believed that this stochastic variability originates in the accretion disks surrounding the supermassive black holes (SMBHs) at the centers of AGN~\citep{peterson2004}. 
However, the physical mechanisms driving this variability are still poorly understood. 

In recent years, modeling the UV/optical light curves of AGN as stochastic processes has shown promise for revealing potential connections between the observed variability and its physical drivers~\citep[e.g.,][]{kelly2009, macleod2010, kelly2011, simm2016, yu2022b}.
In particular, by modeling AGN light curves as a damped random walk (DRW) process, numerous investigations have shown that the best-fit variability amplitude and the characteristic timescale of DRW correlate with both the Eddington ratio (\lledd) and the black hole mass (\mbh)~of AGN~\citep[e.g.,][]{kelly2009, macleod2010, suberlak2021, burke2021, stone2022}. 
The DRW process is the lowest order continuous-time autoregressive moving-average (CARMA) process---CARMA(1,0), which features a broken power-law power spectral density (PSD). 
At low frequencies (i.e., long timescales), the PSD has a power-law exponent of zero, while at high frequencies (short timescales), the PSD has a power-law exponent of $-2$. 
The frequency (timescale) at which the PSD changes exponent is often referred to as the DRW break/damping frequency (timescale).
The magnitude of the best-fit DRW damping timescale ($\sim\,$hundreds of days for a $10^8\,$\msun~SMBH) suggests that the observed stochastic variability might be driven by thermal fluctuations in the accretion disk~\citep[e.g.,][]{kelly2009, sun2020, burke2021}. 

Despite the success of the DRW model, deviations from it have been discovered, primarily at timescales much shorter than the damping timescale of DRW~\citep[e.g.,][]{mushotzky2011, kasliwal2015, smith2018, stone2022, Petrecca2024, Arevalo2024}.
The high cadence and signal-to-noise ratio (S/N) light curves from the {\it Kepler} telescope~\citep{kepler2010} provide the strongest evidence for a deviation~\citep{mushotzky2011, kasliwal2015, smith2018}. 
Alternative models have been proposed/developed to address the deficiency of the DRW description~\citep{kelly2014, kasliwal2017, moreno2019}. 
Notably, \citet[][hereafter \citetalias{kasliwal2017}]{kasliwal2017} demonstrated that a second-order CARMA process, specifically the CARMA(2,1) process, provides a significantly better fit to the {\it Kepler} light curve of ZW229-15. 
The CARMA(2,1) process is also known as the noise-driven damped harmonic oscillator (DHO) model. 
Subsequently, \citet[][hereafter \citetalias{moreno2019}]{moreno2019} performed an extensive study of the variety of PSD shapes that a DHO process can describe.

\citet[][hereafter \citetalias{yu2022b}]{yu2022b} carried out the first attempt to model the light curves of a statistical sample of spectroscopically confirmed quasars as a CARMA(2,1)/DHO process, following the framework prescribed by \citetalias{kasliwal2017} and \citetalias{moreno2019}.
\edit{A DHO process is fully characterized by four independent parameters: the natural oscillation frequency ($\omega_{0}$), the damping ratio ($\xi$), the characteristic timescale of perturbations ($\tau_{\mathrm{perturb}}$), and the amplitude of the driving white noise ($\sigma_{\epsilon}$). The long-term variability amplitude of the process is denoted by $\sigma_{\mathrm{DHO}}$, which is a function of these four parameters.}
\edit{\citetalias{yu2022b} found that \sigmadho~anti-correlates with both \lledd~and \mbh~of the investigated quasars---consistent with results from DRW analyses~\citep[e.g.,][]{macleod2010, suberlak2021}. 
Enabled by the additional freedom of the DHO model, \citetalias{yu2022b} also found that the short-term variability of the best-fit DHO process---as characterized by \sigmanoise~and~\tauperturb---exhibit anti-correlations with quasars' bolometric luminosity (\lbol). }
Due to the limited baseline of the light curves used in \citetalias{yu2022b}, they were not able to probe the correlations of DHO's long-term characteristic timescale (\taudecay, which is similar to the damping timescale of DRW) with quasar physical properties. 
Similarly, due to a limited sample size, they were not able to robustly calibrate the correlations of \sigmanoise~and \tauperturb~with quasar physical properties.

In this work, we apply the DHO model to the 22-year-long $g$- and $r$-band light curves of a large sample of quasars discovered in the Sloan Digital Sky Survey~\citep[SDSS;][]{york2000} Stripe 82 region.
\edit{The light curves are collected and combined from SDSS, the Panoramic Survey Telescope and Rapid Response System~\citep[Pan-STARRS;][]{chambers2016}, and the Zwicky Transient Facility~\citep[ZTF;][]{ZTF2019a, ZTF2019b}, more than tripling the time baseline of the light curves used by~\citetalias{yu2022b}.}
\edit{The much longer temporal baseline of these light curves allows us to extend the investigation conducted in~\citetalias{yu2022b}.}
In particular, we investigate the connection between best-fit DHO parameters and prominent features in their model PSD and structure function (SF). 
We calibrate the correlations of four best-fit DHO parameters (\sigmadho, \taudecay, \sigmanoise, and \tauperturb) with the rest-frame wavelength (\lambdarf) of the observed variable emission.
We also test for correlations between DHO parameters and quasar physical properties, specifically, the monochromatic luminosity at 3000\rmAA~(\lthreezero), the full-width-at-half-maximum of the MgII emission line (\fhmgii), the Eddington ratio (\lledd), and the black hole mass (\mbh).
Lastly, the much extended light curves enable us, for the first time, to assess how empirically inferred DHO parameters depend on the length of the light curve baseline.

The structure of this paper is as follows: in Section~\ref{sec:dho_modeling}, we introduce the fundamentals of a DHO process. 
In Section~\ref{sec:data}, we describe the datasets utilized and the data collection process. 
In Section~\ref{sec:samples}, we describe the light curve fitting process and the steps taken to clean up the resulting sample of DHO fits.
We present the results of our investigation in Section~\ref{sec:results}. 
Finally in Section~\ref{sec:discussion}, we discuss our interpretations for the quasar variability captured by our DHO model, as well as for the correlations between DHO parameters and quasar physical properties. We briefly summarize and conclude in Section~\ref{sec:conclude}.

\section{The DHO Model}\label{sec:dho_modeling}
Here, we briefly describe the key properties of a DHO process. In-depth discussions of this model can be found in \citetalias{kasliwal2017}, \citetalias{moreno2019}, and \citetalias{yu2022b}. 
A DHO process is formally defined as the solution to the following stochastic differential equation (SDE),
\begin{equation}\label{eqn:dho1}
    d^{2} x+\alpha_{1} d^{1} x+\alpha_{2} x=\beta_{0} \epsilon(t)+\beta_{1} d^{1}(\epsilon(t)),
\end{equation}
where $\epsilon(t)$ is Gaussian white noise with an amplitude of unity,\footnote{$\epsilon(t)\equiv dW/dt$, $W$ is the Wiener process or referred to as Brownian motion in physics.} $\alpha_{1}$ and $\alpha_{2}$ are called the autoregressive (AR) coefficients,\footnote{$\alpha_0$ is defined to be 1 by convention.} and $\beta_{0}$ and $\beta_{1}$ are called the moving-average (MA) coefficients~\citep{rouxalet2002, kelly2014,kasliwal2017,moreno2019,yu2022b}.
The differentiation is with respect to time, and $x$ represents the brightness of the quasar.
In an impulse-response dynamical system characterized by Equation~\ref{eqn:dho1}, the left-hand side (LHS) of the SDE describes how the system would respond to a delta-function perturbation and the right-hand side (RHS) specifies how the system is being perturbed/excited.
The PSD of a DHO process is defined as,
\begin{equation}
    {P}_{\rm DHO}(f)=\frac{1}{2\pi}\frac{\beta_0^2+4 \pi^2 \beta_1^2 f^2}{16 \pi^4 f^4+4 \pi^2 f^2\left(\alpha_1^2-2 \alpha_2\right)+\alpha_2^2}
\end{equation}\label{eqn:dho_psd}
where $f$ is the frequency.

Given the close analogy of Equation~\ref{eqn:dho1} to the differential equation of the classical damped harmonic oscillator, Equation~\ref{eqn:dho1} can be re-parameterized to,
\begin{equation}\label{eqn:dho2}
    d^{2} x+2 \xi \omega_{0} d^{1} x+\omega_{0}^{2} x= \sigma_{\mathrm{\epsilon}} \epsilon(t)+\tau_{\mathrm{perturb}}\,\sigma_{\mathrm{\epsilon}}d^{1}(\epsilon(t)),
\end{equation}
where $2\xi\omega_{0} = \alpha_{1}$ and $\omega_{0}^{2} = \alpha_{2}$. On the LHS, $\xi$ is the damping ratio of the damped oscillator, and $\omega_{0}$ is the natural oscillation frequency (i.e., when there is no damping) of the described dynamical system. 
On the RHS, \sigmanoise\ gives the amplitude of the perturbing white noise $\epsilon(t)$, and \tauperturb\ gives a characteristic timescale of the perturbation process. 
Following the classical damped harmonic oscillator, we refer to a DHO as underdamped when $\xi < 1$, and overdamped when $\xi > 1$. 
From the roots of the characteristic equation corresponding to the LHS of Equation~\ref{eqn:dho1}, additional DHO timescales can be derived. 
A complete list of derived DHO timescales/features can be found in \citetalias{moreno2019} and \citetalias{yu2022b}. 
For completeness, we list all DHO-related quantities and their definitions in Appendix~\ref{appendix:dho_feats}.

In this work, we focus on investigating the AGN UV/optical variability captured by four main DHO parameters: the long-term variability amplitude (\sigmadho), the long-term decay timescale (\taudecay), the short-term variability amplitude (\sigmanoise), and the perturbation timescale (\tauperturb).
\sigmadho~is equivalent to the standard deviation of the DHO process---if one could sample it with an infinite duration. 
\taudecay~can be understood as a long-term characteristic timescale beyond which the DHO process becomes de-correlated, i.e., data points separated by more than one \taudecay~follow a white noise PSD.
The main reason for limiting our investigation to only those four properties is that other DHO timescales/features might be poorly constrained (see Section 4 of \citetalias{yu2022b} for an in-depth discussion) even with the 22-year-long light curves utilized here.

\section{Data Sets}\label{sec:data}
The quasars investigated are collected from the SDSS Stripe 82 region~\citep[S82;][]{annis2014}, which is a $120^{\circ}$-long and $2^{\circ}$.5-wide stripe centered along the celestial equator that has been repeatedly imaged by multiple optical surveys including SDSS.
We start with the SDSS Data Release Sixteen Quasar catalog~\citep[DR16Q;][]{lyke2020} and select quasars that are located within the S82 region. 
A total number of $38,846$ quasars are returned.\footnote{\edit{In \citetalias{yu2022b}, a sample of 12,714 quasars was selected from the SDSS DR16Q catalog, excluding objects without spectroscopic measurements reported by \cite{shen2011} or \cite{rankine2020}. In contrast, the sample of 38,846 quasars presented here does not impose any requirement on the availability of spectroscopic properties.}}
We then cross-match this sample of quasars with the archival multi-epoch photometry obtained by SDSS~\citep{york2000}, the Pan-STARRS project~\citep{chambers2016}, and ZTF~\citep{ZTF2019a, ZTF2019b} to construct the light curves, as described below. 
Lastly, we perform spectral energy distribution (SED) fitting to determine the photometric offsets needed to bring Pan-STARRS and ZTF light curves into the SDSS photometric system.
The spectroscopic properties (e.g., emission line properties, luminosity, etc.) of each quasar are adopted from the \citet{wu2022} catalog.
In the rest of this section, we provide more details on each dataset utilized. We note that all cross-matching uses a $1\farcs0$ radius, and PSF photometry is used to construct the light curves.

\subsection{Light Curve Data}\label{subsec:data_lc}

\subsubsection{SDSS (2000--2008)}\label{subsubsec: data_sdss}
SDSS repeatedly imaged the S82 region from 2000 to 2008, producing up to 90 single-epoch observations~\citep{york2000,frieman2008,sako2008}. 
This period spans two phases of SDSS, namely the SDSS Legacy Survey~\citep[2000-2005;][]{york2000} and the SDSS-II Supernova Survey~\citep[2005-2008;][]{frieman2008, sako2008}.
The limiting magnitudes ($5\sigma$) of SDSS, at the 50$\%$ completeness level, are 22.5, 23.2, 22.6, 21.9, and 20.8 in the $u$, $g$, $r$, $i$, and $z$ bands, respectively~\citep{york2000}. 
The SDSS light curves for our quasars are generated by cross-matching against the \texttt{photoobj} table in the \texttt{Stripe82} context on CasJobs.\footnote{\url{https://skyserver.sdss.org/casjobs/}}
The raw light curves consist of matched single-epoch detections whose photometry is marked as ``clean".\footnote{\url{https://www.sdss.org/dr16/tutorials/flags/}} 
We further clean the raw light curves in each band by removing outlier detections that deviate from the seasonal median by more than 5 times the standard deviation of the raw light curve. 
This step removes photometric measurements that display spuriously large deviations from the median (see Figure 2 of~\citealt{schmidt2010}).
\edit{Lastly, we average observations taken within the same night using inverse-variance weighting, to avoid any potential calibration issues.}

\subsubsection{Pan-STARRS1 (2009--2014)}\label{subsubsec: data_ps1}
Pan-STARRS is a wide-field imaging survey developed and operated by the Institute for Astronomy at the University of Hawaii~\citep{chambers2016}. 
The first part of Pan-STARRS (PS1) was conducted from 2009 to 2014. 
The PS1 $3\pi$ Steradian Survey imaged the sky north of Dec $=-\,30^{\circ}$, which fully covers the S82 region, in five bands, $g_{\rm{P1}}, r_{\rm{P1}}, i_{\rm{P1}}, z_{\rm{P1}}$ and $y_{\rm{P1}}$. 
The corresponding $5\sigma$ single-exposure limiting magnitudes are 22.0, 21.8, 21,5, 20.9, and 19.7, respectively. 
Each patch of the sky imaged by the $3\pi$ Survey was observed a total of 60 times on average in all five bands ($\sim\,$12 epochs per band). 
We construct the raw PS1 light curves by first cross-matching our quasars against the \texttt{MeanObject} table, and then collecting the matched PSF photometry from the \texttt{Detection} table while requiring the \texttt{psfQfPerfect} of each photometry to be greater than 0.95. 
\texttt{psfQfPerfect} reports the fraction of totally unmasked (good) pixels for each PSF photometry measurement~\citep{flewelling2020}.
The cross-matching is performed using the MAST Casjobs web portal.\footnote{\url{https://mastweb.stsci.edu/mcasjobs/}}
Lastly, we average observations taken within the same night using inverse-variance weighting.

\subsubsection{ZTF (2017--2023)}\label{subsubsec: data_ztf}
ZTF is an optical time-domain survey that employs a dedicated wide-field camera mounted on the Palomar 48 inch Schmidt telescope~\citep{ZTF2019a, ZTF2019b}.
With a custom-build camera that provides a 47 $\rm deg^2$ field of view and a 8 second readout time, the ZTF pubic program can survey the entire sky north of Dec = $-30^{\circ}$ every $2$--$3$ nights (assuming a 30-second exposure). 
The median single-epoch depth of ZTF reaches $\sim21.2$, $\sim21$, and $\sim20.5$ in $g$, $r$, and $i$ bands, respectively.
To compile the ZTF light curves, we cross-match our quasars with the Zubercal\footnote{\url{http://atua.caltech.edu/ZTF/Zubercal.html}} version of ZTF Data Release 20. 
\edit{Zubercal is a re-calibrated photometric dataset based on ZTF's science-image-based PSF photometry, offering improved photometric accuracy, proper color corrections, and more realistic uncertainty estimates compared to the original data release photometry.\footnote{\url{https://irsa.ipac.caltech.edu/data/ZTF/zubercal/overview.pdf}}}
Given the matched Zubercal light curves, we first remove outlier photometric measurements that deviate from the seasonal median by more than 5 times the standard deviation ($\sigma$) of the raw light curve. 
\edit{We then average observations taken within the same night using inverse-variance weighting.} 
The adopted 5$\sigma$ cut follows the same motivation as that applied to the SDSS light curves---to remove photometric measurements that show spurious deviations from the median. 
We note that the median 5$\sigma$ value of our quasar light curves is $\sim1$ magnitude, which is much larger than the expected amplitude of quasar stochastic variability at the few-month timescale~\citep{vandenberk2004}, and thus the adopted 5$\sigma$ cut removes mostly outlier data points rather than intrinsic quasar variability.

\subsection{Validation \& Recalibration of Photometric Errors}\label{subsec:err-calib}

\begin{figure}
    \centering
    \includegraphics[width=0.99\linewidth]{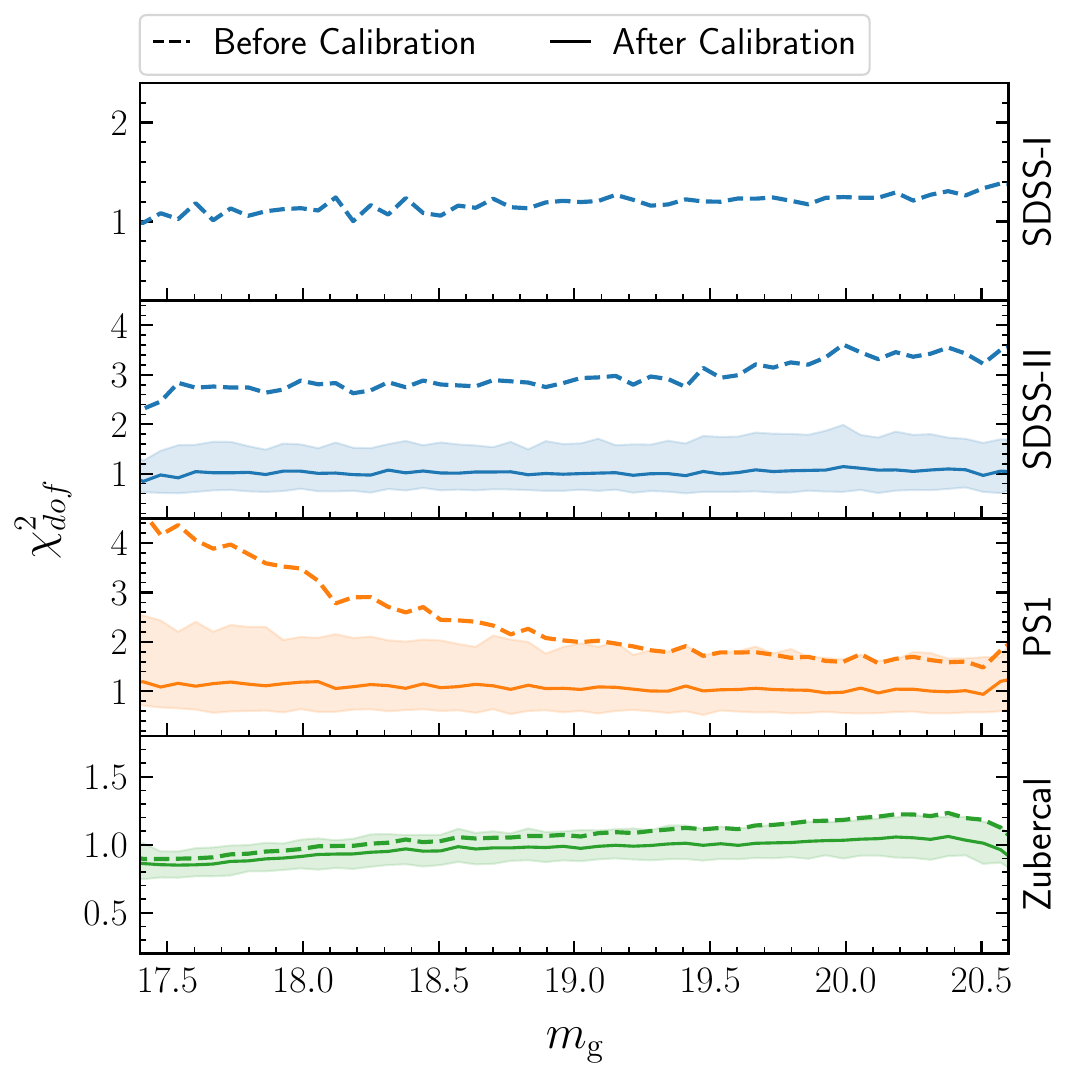}
    \caption{$\chi^2$ per degree of freedom ($\chi^2_{dof}$) distribution for standard star light curves compiled from the SDSS-I Legacy Survey, the SDSS-II Supernova Survey, PS1, and ZTF~(Zubercal). The dashed lines show the median of the $\chi^2$ distribution computed using the pipeline photometric errors, as a function of the star's mean magnitude. The solid lines and the shaded regions show the $\chi^2$ distribution computed using the re-calibrated photometric errors. Recalibration is not performed for SDSS-I photometry.}
    \label{fig:err_calib}
\end{figure}

We use $\sim21,000$ randomly selected sources from the Stripe 82 Standard Star catalog~\citep{Thanjavur2021} to validate the pipeline photometric error of the photometry in our light curves. 
We follow the same steps as described in Section~\ref{subsubsec: data_sdss},~\ref{subsubsec: data_ps1}, and~\ref{subsubsec: data_ztf} to compile the SDSS, PS1, and Zubercal light curves of these standard stars. 
We then compute the $\chi^2$ per degree of freedom statistic for each light curve. 
If the photometric errors are perfectly calibrated, the distribution of $\chi^2$ should be centered at unity~\citep{ivezic2007}. In contrast, $\chi^2$ should be greater than unity for underestimated errors, and less than unity for overestimated errors.
The dashed lines in Figure~\ref{fig:err_calib} show the median of the $\chi^2$ distribution computed using the pipeline errors. 
The photometric error of SDSS-I photometry appears to be well-calibrated, except for a slight underestimation at the faint end. 
The photometric errors of SDSS-II and PS1 photometry are noticeably underestimated.
The photometric error of Zubercal photometry also appears well-calibrated, aside from a slight underestimation at the faint end. 

We derive a corrective procedure for each survey's pipeline photometric error by fitting their standard star light curves with a Gaussian process (GP) model. This GP model has a specific kernel:
\begin{equation}
    k_{nm} = \sigma_{\rm excess}^2\,\delta_{nm} + \sigma_{n}^2\,\delta_{nm}
\end{equation}
where $\sigma_{\rm excess}$ is defined as the amplitude of the GP kernel, which models the excess variance of the light curve that is not accounted for by the pipeline photometric error---$\sigma_{\rm n}$. 
Next, we fit $\sigma_{\rm excess}$ as a function of the star's mean magnitude using a third-order polynomial. Finally, we add $\sigma_{\rm excess}(m)$, where $m$ is the magnitude of the photometric measurement, to the pipeline photometric error in quadrature.
The solid lines in Figure~\ref{fig:err_calib} show the median of the $\chi^2$ distribution computed using the re-calibrated photometric errors, and the shaded regions show the corresponding 1$\sigma$ range (the central 68$\%$ interval). 
We do not perform error recalibration for SDSS-I photometry, because the best-fit $\sigma_{\rm excess}$ for more than half of the standard stars are zero, suggesting that the pipeline uncertainties are already well-calibrated.
This corrective procedure is later applied to the pipeline errors of the photometry in our quasar light curves.

\subsection{SED Fitting \& Light Curve Merging}\label{subsec:data_lc_merge}
We create combined SDSS-PS1-ZTF light curves for our quasars by applying photometric offsets to PS1 and ZTF photometry.
The photometric offset for each quasar is determined using synthetic SDSS and PS1 photometry computed from the model spectrum produced by the SED-fitting code \texttt{qsogen}\footnote{\url{https://github.com/MJTemple/qsogen}}~\citep{temple2021}.
To perform the SED fitting, we collect coadded $g$, $r$, $i$, and $z$ band photometry from the SDSS Equatorial Survey of the S82 region~\citep{annis2014}.
The coadded SDSS photometry were created from SDSS images taken from 1998 to 2005, reaching limiting magnitudes that are $\sim\,$2 mag fainter than the SDSS single-pass data. 
Given that SDSS images used in the co-addition span a long baseline (7 years) and that SDSS obtains observations in different bands nearly simultaneously, we argue that the coadded multi-band photometry provides a reasonable representation of each quasar's average SED.

In addition to a model spectrum, \texttt{qsogen} also provides a best-fit \lthreezero.
We update the \lthreezero~measurement from \citet{wu2022} with the \texttt{qsogen} estimate.
We note that our main results will not change if we use the \lthreezero~directly measured from the SDSS spectrum; however, in the interest of consistency, we choose to use the \lthreezero~estimate provided by \texttt{qsogen}.


\subsection{The Initial Sample}\label{subsec:main_sample}
Our \texttt{initial sample} consists of $21,767$ unique quasars that meet the following primary criteria:
\begin{itemize}
    \item The quasar is detected by all three surveys (i.e., SDSS, PS1, and ZTF) in both $g$- and $r$-band. 
    \item The mean SDSS magnitude (computed from the light curve) of the quasar is $<$ 22.0 in the $g$ band and $<$ 21.8 in the $r$ band. \edit{This cut ensures that the matched ZTF and PS1 light curves do not predominantly sample the bright phase of the given quasar, which could arise from the shallower depths of ZTF and PS1 relative to SDSS.}
    \item The quasar has high S/N measurements of its spectroscopic properties. Specifically, we require the uncertainty of \fhmgii~to be $<$ 10,000 $\rm km\,s^{-1}$, the S/N of \fhmgii~to be $>$ 2, and the S/N of the Mg\,\textsc{ii} broad component luminosity to be $>$ 2. 
    This criterion effectively selects quasars between $z\simeq0.28$ and $z\simeq2.71$. 
\end{itemize}
After the first, second, and third cut, we are left with $31,626$, $27,050$, and $22,027$ unique quasars, respectively. 
Note that we only consider $g$- and $r$-band light curves, because the archival ZTF $i$-band light curves are significantly shorter than their $g$- and $r$-band light curves. 
The limited $i$-band baseline could result in biased DHO fits in comparison to the $g$- and $r$-band DHO fits, especially for the long-term variability parameters (\sigmadho~and~\taudecay).
\edit{We further restrict our analysis to quasars with high S/N spectroscopic measurements of the Mg\,\textsc{ii} line, since high S/N spectroscopic measurement of other physical properties---such as the H$\beta$ line in low-$z$ quasars and the C\,\textsc{iv} line in high-$z$ quasars---are available for significantly fewer sources.}
Lastly, we require each quasar to have a well-constrained \texttt{qsogen} SED model, i.e., the returned covariance matrix for the fitted parameters does not contain infinity. This cut ensures that there is a SED-based photometric offset to calibrate photometry from ZTF/PS1 to SDSS, and it removes an additional $260$ quasars.
\edit{After applying these selection criteria, we are left with 21,767 unique quasars, which constitute our \texttt{initial sample}.}

\begin{figure}
    \vspace{0.2cm}
    \centering
    \hspace{-0.1cm}\includegraphics[width=0.98\linewidth]{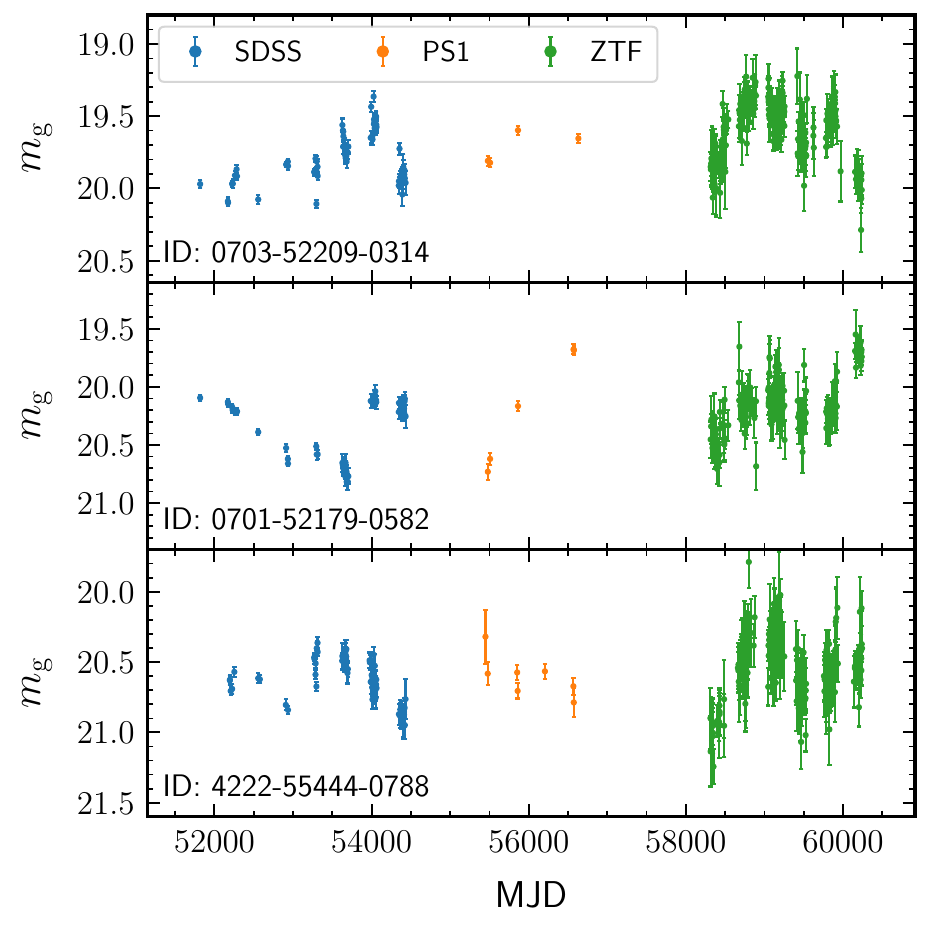}
    \caption{Three example combined SDSS-PS1-ZTF light curves. 
    The \texttt{ID} shown at the bottom left corner of each panel gives the identifier of each corresponding quasar's SDSS spectrum, in the format of \texttt{PLATE}-\texttt{MJD}-\texttt{FIBERID}.}
    \label{fig:example_lc}
\end{figure}

\edit{The SDSS-PS1-ZTF combined light curves of quasars in the \texttt{initial sample} span a median observed-frame baseline of 22 years and contain a minimum of 28 epochs. 
The average temporal gap between the SDSS and PS1 light curves is $\sim$1050 days, while the gap between PS1 and ZTF averages $\sim$1750 days.
Table~\ref{tab:lcStat} summarizes the cadence statistics of the combined $g$-band light curves, as well as those from the individual surveys.}
Three example $g$-band light curves are shown in Figure~\ref{fig:example_lc}.
The \texttt{ID} at the bottom left corner of each panel has the format of \texttt{PLATE}-\texttt{MJD}-\texttt{FIBERID}, where \texttt{PLATE}, \texttt{MJD}, and \texttt{FIBERID} together identify the SDSS spectrum of the corresponding quasar.

\begin{deluxetable*}{lccccc}
    \setlength{\tabcolsep}{10pt}
    \tablecaption{Cadence statistics for the $g$-band light curves}
    \tablehead{
    \colhead{Cadence Statistic}
    &\colhead{Sample}
    &\colhead{SDSS}
    &\colhead{PS1}
    &\colhead{ZTF}
    &\colhead{SDSS+PS1+ZTF}
    }
    \startdata
    16th Percentile $\Delta t$ [day]    & Initial & $2.0^{-0.0}_{+0.0}$ & $22.0^{-16.2}_{+159.5}$ & $1.9^{-0.9}_{+1.0}$ & $2.0^{-1.0}_{+0.0}$\\
                                        & Clean   & $2.0^{-0.0}_{+0.0}$ & $23.7^{-18.0}_{+158.1}$ & $1.7^{-0.7}_{+1.2}$ & $2.0^{-0.9}_{+0.0}$\\
                                        & Core    & $2.0^{-0.0}_{+0.0}$ & $35.2^{-28.6}_{+146.7}$ & $1.4^{-0.4}_{+1.3}$ & $2.0^{-0.9}_{+0.0}$\\[.5ex]
    \hline
    Median $\Delta t$ [day]     & Initial & $5.0^{-1.0}_{+1.0}$ & $346.0^{-287.3}_{+11.1}$ & $3.9^{-1.0}_{+9.0}$ & $4.0^{-1.0}_{+3.0}$\\
                                & Clean   & $5.0^{-1.0}_{+1.0}$ & $347.0^{-181.7}_{+10.6}$ & $3.9^{-1.0}_{+9.0}$ & $4.0^{-1.0}_{+3.0}$\\
                                & Core    & $5.0^{-1.0}_{+1.0}$ & $346.0^{-172.0}_{+10.6}$ & $3.5^{-0.6}_{+8.7}$ & $4.0^{-1.0}_{+3.0}$\\[.5ex]
    \hline    
    84th Percentile $\Delta t$ [day]    & Initial & $20.6^{-4.8}_{+5.7}$ & $365.6^{-13.5}_{+182.6}$ & $17.9^{-9.9}_{+37.7}$ & $21.9^{-11.8}_{+65.8}$\\
                                        & Clean   & $19.5^{-4.6}_{+6.1}$ & $366.4^{-13.6}_{+184.7}$ & $17.8^{-9.8}_{+40.1}$ & $20.9^{-10.9}_{+43.6}$\\
                                        & Core    & $19.0^{-4.4}_{+6.3}$ & $366.5^{-12.7}_{+187.4}$ & $20.0^{-9.0}_{+38.9}$ & $20.9^{-10.0}_{+41.4}$\\[.5ex]
    \hline
    Baseline [day]              & Initial & $2236.9^{-335.1}_{+367.0}$ & $1097.0^{-362.0}_{+32.0}$ & $1919.8^{-735.0}_{+40.0}$ & $8040.0^{-455.2}_{+353.1}$\\
                                & Clean   & $2236.9^{-38.0}_{+378.0}$  & $1097.0^{-369.0}_{+32.0}$ & $1919.8^{-411.4}_{+38.0}$ & $8042.0^{-353.9}_{+377.0}$\\
                                & Core    & $2236.9^{-28.9}_{+378.0}$  & $1097.0^{-369.0}_{+32.0}$ & $1919.8^{-394.2}_{+38.0}$ & $8042.0^{-360.0}_{+378.0}$\\[.5ex]
    \hline
    Number of Epochs            & Initial &  $57^{-10}_{+10}$  &  $5^{-1}_{+2}$ & $108^{-86}_{+130}$ & $173^{-92}_{+132}$\\
                                & Clean   &  $60^{-10}_{+11}$  &  $5^{-1}_{+2}$ & $111^{-85}_{+126}$ & $181^{-93}_{+116}$\\
                                & Core    &  $60^{-11}_{+10}$  &  $5^{-1}_{+2}$ & $112^{-84}_{+118}$ & $184^{-93}_{+121}$\\    
    \enddata
    \tablecomments{Each value reports the 16th, 50th (median), and 84th percentiles of the cadence statistic distribution, displayed as subscript, central value, and superscript, respectively. 
    $\Delta t$ denotes the time interval between adjacent observations within a given light curve.}
\end{deluxetable*}\label{tab:lcStat}


\section{Light Curve Fitting \& Sample Cleaning}\label{sec:samples}

\begin{deluxetable}{ccc}
    \setlength{\tabcolsep}{4pt}
    \tablecaption{Fitting process configuration}
    \tablehead{\colhead{DHO Parameter} & \colhead{Initial Guess} & \colhead{Best-fit Prior}}
    \startdata
    log($\alpha_1$)  & Uniform[-2.6, 2.6]   & Uniform[-6.5, 6.5]   \\
    log($\alpha_1$)  & Uniform[-2.6, 2.6]   & Uniform[-6.5, 6.5]   \\
    log($\beta_0$)   & Uniform[-4.4, 0.87]  & Uniform[-10, 3]      \\
    log($\beta_1$)   & Uniform[-4.4, 0.87]  & Uniform[-10, 3]      \\
    \enddata
    \tablecomments{The DHO parameter range for initial guesses corresponds to a maximum \taudecay~of $\sim10^5$ days and a minimum \tauperturb~of $\sim10^{-1}$ day.}
    \vspace{-1cm}
\end{deluxetable}\label{tab:fit_prior}

\subsection{Light Curve Fitting}\label{subsec:lc_fit}
We follow the approach developed in \citetalias{yu2022b} to fit a DHO model to the observed-frame magnitude light curves of \texttt{initial sample} quasars.
We emphasize that fitting a DHO (or CARMA) model to magnitude light curves rather than flux light curves is a reasonable choice, because a CARMA process is Gaussian by definition~\citep{rouxalet2002, brockwell2001, kelly2014}, while the distribution of magnitude light curves by accreting compact objects is also Gaussian~\citep{uttley2001, gaskell2004, macleod2012, decicco2022}. 
During the fitting process, we start by randomly initializing 300 optimizers across the plausible DHO space set by the temporal sampling of the input light curve, and take the maximum a posteriori (MAP) estimate as the `best-fit' solution. 
The DHO parameter ranges used in optimizer initialization (initial guess) and the flat priors used in the best-fit search are summarized in Table~\ref{tab:fit_prior}.
\edit{These choices are informed by the cadence characteristics of our light curves.}
Once the best-fit DHO parameters are obtained, we use \texttt{emcee}~\citep{foreman-mackey2013}, a python implementation of Goodman \& Weare's~(\citeyear{goodman2010}) Affine Invariant Markov chain Monte Carlo (MCMC) Ensemble sampler, to sample the posterior distribution. 
The MCMC priors are the same as those used in the best-fit search.
We initiate a total of 32 walkers at the MAP positions, and run the MCMC walkers for 40,000 steps. 
We discard the first 10,000 steps as the ``burn-in" phase.
We adopt the 1$\sigma$ range (the central 68\% interval) of the marginalized posterior distribution as the uncertainty of each best-fit DHO parameter. 
Note that we run MCMC in the log space of ${\alpha_1}$, ${\alpha_2}$, ${\beta_0}$, and ${\beta_1}$, and thus the reported uncertainties of DHO parameters are log quantities.

\subsection{The Clean Sample}\label{subsec:clean_sample}
Given the best-fit DHO parameters and their uncertainties, we clean our \texttt{initial sample} DHO fits using three main criteria.
First, we remove DHO fits that have either a \taudecay~longer than half the light curve baseline, or with a \tauperturb~shorter than twice the minimum separation between any two observations. 
These removed DHO fits have derived timescales that are nearly impossible to reliably constrain given the cadence and length of their light curves~\citep[e.g.,][]{kozlowski2017a, burke2021}.
The \taudecay-based cut alone removes $\sim15\%$ of the fits, and the \tauperturb-based cut alone removes $\sim40\%$ of the fits.
Second, we remove DHO fits with log($\xi$) $-$ log($\omega_{0}$) $>$ 1, which corresponds to a region in the DHO space where the daily observing cadence could dominate over the intrinsic variability signal~(see Section 4.1 of \citetalias{yu2022b}).
This cut alone removes $\sim10\%$ of the fits.
After the cuts described above, we are left with $10,435$ $g$-band DHO fits and $8,907$ $r$-band DHO fits.
Lastly, we remove DHO fits that have a posterior distribution that appears bi- or multi-modal. Specifically, we remove DHO fits with $|{\rm log}(X_{\rm MAP})- {\rm log}(X_{\rm median})| > 0.9$, where $X$ is one of $\alpha_1$, $\alpha_2$, and \tauperturb. $X_{\rm MAP}$ is the MAP estimation of $X$, and $X_{\rm median}$ is the median of the marginalized MCMC distribution of $X$. 
We also remove DHO fits with $\sigma_{\alpha_1}+\sigma_{\alpha_2}+\sigma_{\tau_{perturb}} < 3$, where $\sigma_{\alpha_1}$, $\sigma_{\alpha_2}$, and $\sigma_{\tau_{perturb}}$ are the MCMC-based uncertainties of $\alpha_1$, $\alpha_2$, and \tauperturb, respectively.
The limits for the MCMC-based cuts are chosen based on simulations performed to validate our sample cleaning procedures~(see Section~\ref{subsec:sim_fit} for more details of the simulation carried out).
Our final \texttt{clean sample} contains $2,888$ $g$-band DHO fits and $1,861$ $r$-band DHO fits, \edit{with 422 quasars having both $g$- and $r$-band fits}.


\begin{figure*}
    \centering
    \includegraphics[width=0.9\linewidth]{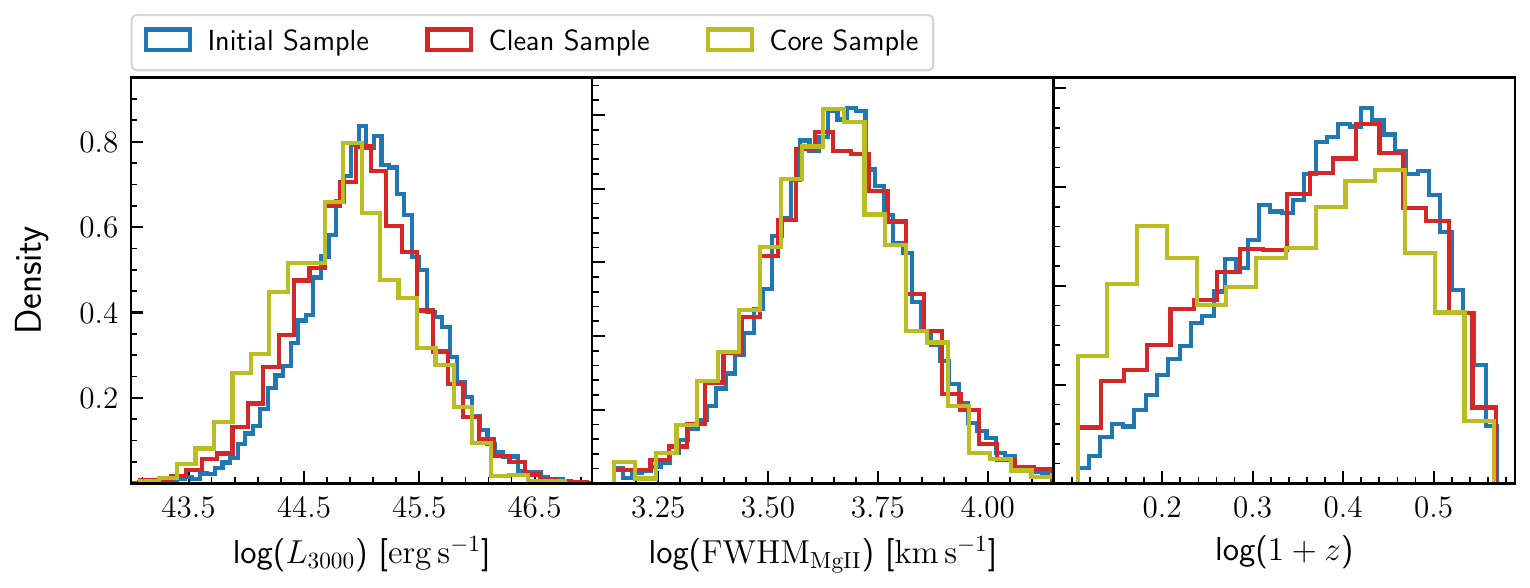}
    \caption{The distributions of \texttt{initial sample}, \texttt{clean sample}, and \texttt{core sample} quasars in \lthreezero, \fhmgii, and log($1+z$).  
    \edit{The \texttt{clean sample} and \texttt{core sample} quasars are centered at slightly lower \lthreezero~than those in the \texttt{initial sample}, but their \fhmgii~distributions appear comparable.
    The modest shift in \lthreezero~for the \texttt{clean sample} and \texttt{core sample} can be attributed to the \taudecay-based cut, which preferentially excludes high-redshift quasars (and thus higher-luminosity sources) due to the redshift-dependent contraction of the rest-frame baseline. This effect is also illustrated in the $\log(1+z)$ distributions.}
    }
    \label{fig:sample_dist}
\end{figure*}

\subsection{The Core Sample}\label{subsec:core_sample}
From the \texttt{clean sample}, we make additional cuts to create the \texttt{core sample}, which are used to determine the correlations of DHO parameters with \lambdarf, \lthreezero, and \fhmgii.
We first remove DHO fits with a \taudecay~greater than $25\%$ of the light curve baseline; this cut is motivated by the findings that the DRW damping timescale cannot be robustly constrained if the light curve baseline is not at least 10 times longer~\citep{kozlowski2017a, sanchez2017}.
We note that this \taudecay-based cut only affects the results shown in Section~\ref{subsec:phys_corr} (i.e., the coefficients of Equation~\ref{eqn:propCorr} and~\ref{eqn:propCorr2}), and our results will be nearly unchanged if we tighten this cut to $10\%$ of the baseline.
\edit{This cut removes $\sim$22\% of the $g$-band fits and $\sim$25\% of the $r$-band fits from the \texttt{clean sample}. The remaining sample comprises 339 quasars with both $g$- and $r$-band fits.}

\edit{We define the \texttt{core sample} by selecting pairs of $g$-band and $r$-band fits whose parent quasars have similar values in log(\lthreezero), log(\fhmgii), and log($1+z$), where $z$ denotes redshift.
This selection process increases the number of matched $g$-band and $r$-band fits available for investigating correlations between DHO parameters and quasar physical properties, under the assumption that quasars with comparable physical properties exhibit similar temporal variability in luminosity. } More specifically, we match every $g$-band fit with a $r$-band fit while requiring their parent quasars to fall within 0.07 dex in the log(\lthreezero)-log(\fhmgii)~space, where 0.07 dex is the median uncertainty of \texttt{clean sample} quasars' log(\lthreezero) and log(\fhmgii) added in quadrature. 
We further require the parent quasars of each pair of $g$-band and $r$-band fits to be closer than 0.02 in log($1+z$).
If a matching $r$-band fit is not found, the corresponding $g$-band is not selected.
Our \texttt{core sample} contains \edit{1281} $g$-band and \edit{1281} $r$-band fits with matching physical properties and redshift for the quasars.


\subsection{Justification for the Cleaning Cuts}\label{subsec:sim_fit}
We perform simulations to validate our cleaning procedures described in Section~\ref{subsec:clean_sample}.
We suspect that the large number of bad DHO fits removed in Section~\ref{subsec:clean_sample} is a result of the poor light curve quality. If so, DHO fits obtained on DHO light curves simulated following the actual SDSS-PS1-ZTF cadence and photometric S/N should also contain a large number of bad fits as defined by the cuts used in Section~\ref{subsec:clean_sample}.
We simulate $\sim20,000$ $g$-band and $r$-band DHO light curves with their input parameters drawn from the distribution of \texttt{clean sample} DHO fits.
The temporal sampling and photometric S/N of each simulated light curve is set by a randomly chosen quasar light curve from the \texttt{initial sample}.
We fit the DHO model to each realistically-simulated DHO light curve, and clean the best fits using the first two criteria described in Section~\ref{subsec:clean_sample}.
As a result, $\sim35\%$ of the fits are removed; in comparison, these same criteria remove $\sim55\%$ of our \texttt{initial sample} DHO fits.
Next, we run MCMC for the DHO fits that pass these cuts.
We find that ${\rm log}(X_{\rm MAP}) - {\rm log}(X_{\rm median})$ centers at zero with a standard deviation of $\sim0.3$ for each of $\alpha_1$, $\alpha_2$, and \tauperturb.
We also find that the distribution of $\sigma_{\alpha_1}+\sigma_{\alpha_2}+\sigma_{\tau_{perturb}}$ drops significantly beyond 3. 
We remove DHO fits with $|{\rm log}(X_{\rm MAP}) - {\rm log}(X_{\rm median})| > 0.9$~(i.e., three times its standard deviation) and $\sigma_{\alpha_1}+\sigma_{\alpha_2}+\sigma_{\tau_{perturb}} > 3$.
In the end, $\sim33\%$ of the simulated light curves result in clean DHO fits. 
In comparison, $\sim13\%$ of our \texttt{initial sample} light curves have clean DHO fits, which constitute the \texttt{clean sample}.

From these simulation results, we conclude that our cleaning procedures indeed remove unreliable DHO fits. We notice that the same cuts select more clean DHO fits from the simulated sample than from our \texttt{initial sample} ($33\%$ vs.~$13\%$); we propose that the difference can be attributed to \texttt{initial sample} quasars that have intrinsic timescales falling outside the temporal coverage of our light curves (e.g., with an intrinsic \taudecay~longer than the light curve baseline).

\edit{Furthermore, the \texttt{clean sample} is not appreciably skewed relative to the \texttt{initial sample} in terms of light curve characteristics or quasar physical properties. 
Table~\ref{tab:lcStat} lists the cadence statistics for the combined light curves, as well as those from individual surveys. The distributions of these cadence statistics for the \texttt{clean sample} are comparable to those of the \texttt{initial sample}. 
Figure~\ref{fig:sample_dist} shows the distributions of the \texttt{initial sample}, \texttt{clean sample}, and \texttt{core sample} quasars in \lthreezero, \fhmgii, and log($1+z$). 
The \texttt{clean sample} and \texttt{core sample} quasars are centered at slightly lower \lthreezero~than those in the \texttt{initial sample} (left panel in Figure~\ref{fig:sample_dist}), but their \fhmgii~distributions appear comparable (middle panel).
Kolmogorov-Smirnov tests yield $p$-values below 0.05 for comparisons of the \lthreezero~distributions between these samples, whereas the corresponding $p$-values for the \fhmgii~distributions are well above 0.05. 
We attribute the modest shift in \lthreezero~for the \texttt{clean sample} and \texttt{core sample} to the \taudecay-based cut, which preferentially excludes high-redshift quasars (and thus higher-luminosity sources) due to the redshift-dependent contraction of the rest-frame baseline. This effect is also illustrated in the $\log(1+z)$ distributions shown in the right panel of Figure~\ref{fig:sample_dist}.}

\subsection{Best-fit DHO Parameters vs. Light Curve Baselines}\label{subsec:dho_lc_length}

\begin{figure}
    \vspace{0.2cm}
    \centering
    \hspace{-0.1cm}\includegraphics[width=0.95\linewidth]{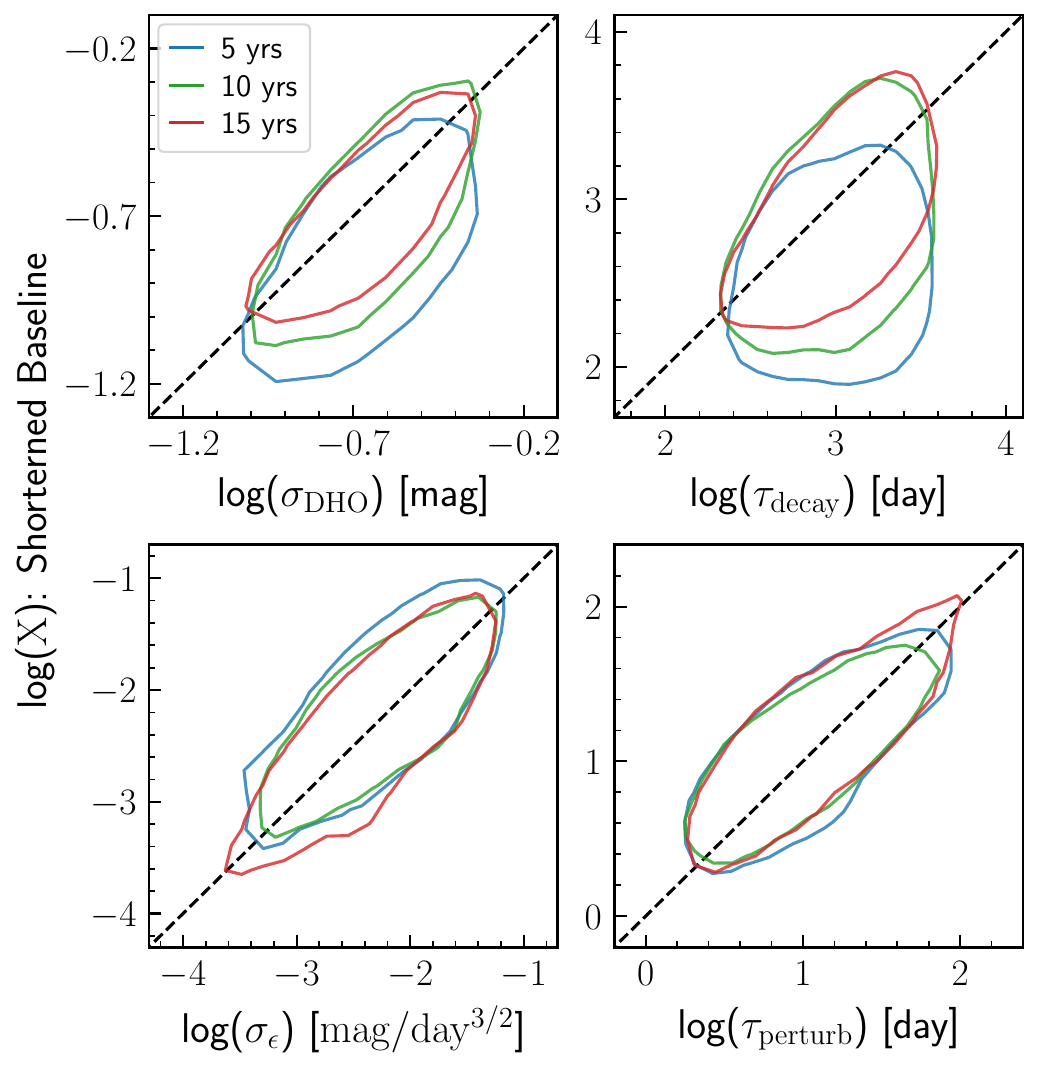}
    \caption{A comparison between the DHO parameters inferred from the full light curves with a 22-year baseline (x-axis) and those inferred from light curves with shortened baselines (y-axis). 
    The blue, green, and red contours show the distributions (80 percentile) of DHO parameters inferred from light curves with 5-year, 10-year, and 15-year baselines, respectively.
    The inferred \sigmadho~and \taudecay~decrease systematically with a decreasing baseline, while \sigmanoise~and \tauperturb~do not show systematic shifts with a decreasing baseline.
    }
    \label{fig:baseline}
\end{figure}

We investigate how the inferred DHO parameters depend on the baseline of the input light curve.
We chop our \texttt{clean sample} light curves with an average baseline of 22 years into segments of 5-year, 10-year, and 15-year baselines; for each baseline, we construct light curves from two temporal ranges.
For the 5-year baseline light curves, the two temporal ranges are $[2003, 2008]$ and $[2018, 2023]$.
For the 10-year baseline light curves, the two temporal ranges are $[2001, 2011]$ and $[2005, 2015]$.
For the 15-year baseline light curves, the two temporal ranges are $[2000, 2015]$ and $[2005, 2020]$.
We then refit the shortened light curves using the method described in Section~\ref{subsec:lc_fit}, and select clean DHO fits using the criteria described in Section~\ref{subsec:clean_sample}. 
We compare DHO parameters inferred from the shortened light curves with those inferred from the full light curves, and plot their distributions in Figure~\ref{fig:baseline}.
Figure~\ref{fig:baseline} shows that the inferred \sigmadho~and \taudecay~of our \texttt{clean sample} quasars decrease systematically when the light curve baseline decreases. 
In contrast, \sigmanoise~and \tauperturb~do not show clear systematic shifts with a decreasing baseline. 
These trends suggest that robust inferences of \sigmanoise~and \tauperturb~are less dependent on super-long light curves in comparison to \sigmadho~and \taudecay.


\begin{figure*}
    \hspace{.3cm}\includegraphics[width=.95\linewidth]{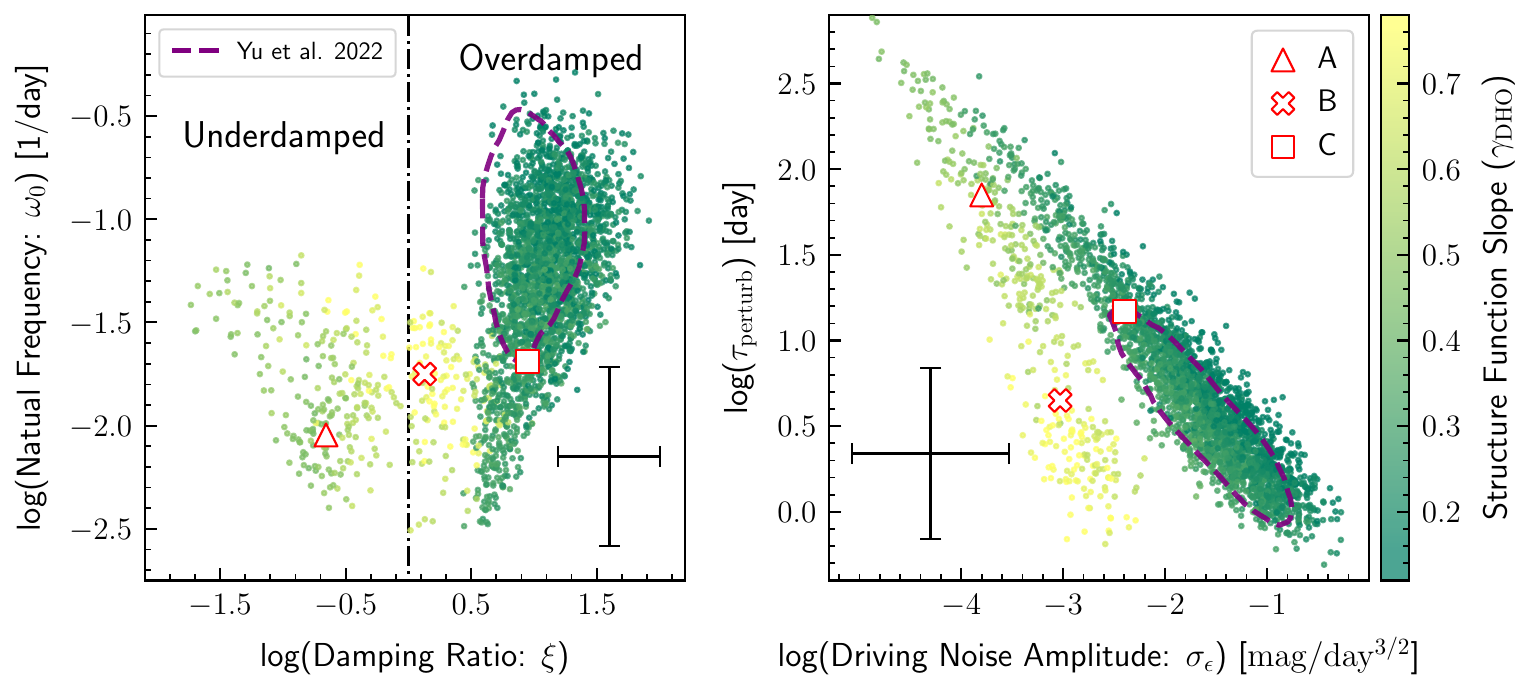}
    \hspace{.0cm}\includegraphics[width=.95\linewidth]{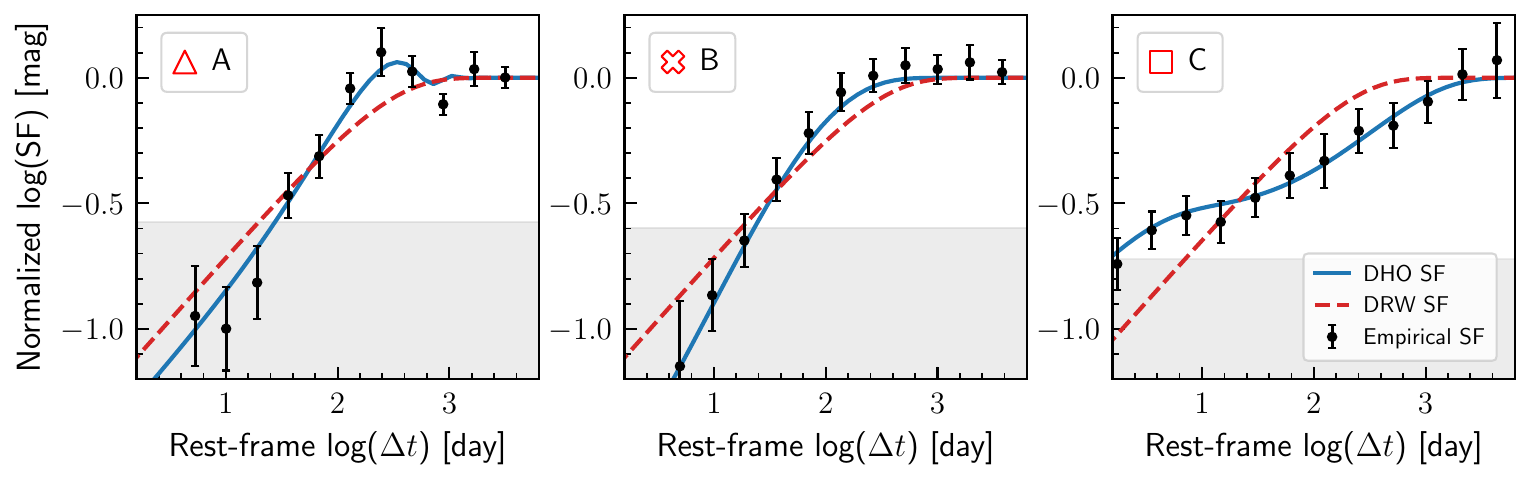}
    \caption{\textit{Top}: The rest-frame distribution of $g$-band DHO fits in the \texttt{\edit{clean} sample}. 
    The left panel shows the DHO space of $\xi$ and $\omega_0$ (LHS of Equation~\ref{eqn:dho2}) and the right panel shows the DHO space of \sigmanoise~and \tauperturb~(RHS of Equation~\ref{eqn:dho2}), where the color gives the SF slope of each best-fit DHO, as defined by Equation~\ref{eqn:sf_slope}. 
    The vertical dash-dotted line in the left panel divides the parameter space into underdamped and overdamped regions.
    Underdamped DHOs have a SF slope steeper than that of a DRW (i.e., 0.5); overdamped DHOs with log($\xi) \lesssim0.5$ have a SF slope steeper than 0.5, while overdamped DHOs with log($\xi) \gtrsim0.5$ have a SF slope flatter than 0.5.
    The median error bars of best-fit DHO parameters are plotted at the bottom-right corner of the left panel and the bottom-left corner of the right panel. 
    The purple contour shows the distribution of DHO fits in the final overdamped sample of \citetalias{yu2022b}.
    {\it Bottom}: The DHO model SF (blue solid line), the comparison best-fit DRW SF (red dashed line), and the empirical ensemble SF (black point) for three representative points marked as \textbf{A}, \textbf{B}, and \textbf{C} in the top panels. 
    All SFs are normalized to a maximum value of 1.
    The grey region indicates the noise floor for the input light curve.
    The DHO SFs follow the empirical SFs better than the DRW SFs. 
    }
    \label{fig:dho_dist}
\end{figure*}

\section{Results}\label{sec:results}
In this section, we examine the diverse quasar variability captured by the DHO process using DHO model SF and model PSD.
In Section~\ref{subsec:dho_dist}, we present the distribution of \texttt{clean sample} DHO fits, and show how the SF shape changes across the DHO parameter space. 
In Section~\ref{subsec:flat_vs_steep}, we explore the relationship between DHO parameters and prominent features in DHO model PSDs.
In Section~\ref{subsec:phys_corr}, we determine the correlations between DHO parameters and the physical properties of quasars. 
In Section~\ref{subsec:vs_yu22}, we highlight how the results of this work compare to those presented in \citetalias{yu2022b}.
The best-fit DHO parameters reported from hereafter are all rest-frame values, which are derived using their scaling relations with $z$ reported by \citetalias{yu2022b}. 
In particular, \sigmadho~does not scale with redshift, \tauperturb~and \taudecay~scale with redshift as $(1+z)^{-1}$, and \sigmanoise~scales with redshift as $(1+z)^{3/2}$.

\subsection{Best-fit DHO Parameters vs. SF Shapes}\label{subsec:dho_dist}
We first investigate the relationship between DHO parameters and DHO model SF shapes~(see Appendix~\ref{appendix:dho_feats} for a formal definition of DHO SF).
Empirically, the SF shape of a typical quasar light curve is often described using a power-law:
\begin{equation}
    {\rm SF(\Delta t)} = {\rm A}\,(\frac{\Delta t}{\Delta t_{0}})^{\gamma}
\end{equation}
where $\Delta t_{0}$ is a characteristic timescale, and ${\rm A}$ is the normalization of the SF at $\Delta t = \Delta t_0$~\citep[e.g.,][]{vandenberk2004, schmidt2010}. Hereafter, we refer to the exponent of the power-law ($\gamma$) as the SF slope. 
The rest-frame distribution of our \texttt{clean sample} $g$-band DHO fits is shown by the top two panels of Figure~\ref{fig:dho_dist}. 
The vertical dash-dotted line in the top-left panel divides the $\xi$--$\omega_0$ parameter space into underdamped DHOs and overdamped DHOs. 
The color-coding gives the SF slope of the best-fit DHO model.
Since DHO SF does not have a fixed shape~(see Figure 16 of \citetalias{moreno2019} for the variety SF shapes can be described by a DHO), we define the SF slope of a DHO as,
\begin{equation}\label{eqn:sf_slope}
    \gamma_{\rm DHO} = \frac{{\rm log}[\rm SF_{DHO}(\tau_{\rm decay})] - {\rm log}[\rm SF_{DHO}(1~day)]}{\rm log(\tau_{\rm decay}) - log(1~day)}
\end{equation}
where $\rm SF_{DHO}(\tau_{\rm decay})$ and $\rm SF_{DHO}(1~day)$ are the DHO SF values at \taudecay~and 1 day, respectively.
Underdamped DHOs have a SF slope steeper than that of a DRW (i.e., 0.5); overdamped DHOs with log($\xi$) $\lesssim0.5$ have a SF slope steeper than 0.5, while overdamped DHOs with log($\xi$) $\gtrsim0.5$ have a SF slope flatter than 0.5.
\edit{As shown in the top-right panel of Figure~\ref{fig:dho_dist}, the SF slope appears to decrease with increasing log(\sigmanoise) and log(\tauperturb).}

To better illustrate how the captured variability changes across the DHO parameter space, we select three points from the distribution of clean DHO fits, and compare their model DHO SFs.
The three points---\textbf{A} (red triangle), \textbf{B} (red cross), and \textbf{C} (red square) are shown in the top two panels of Figure~\ref{fig:dho_dist}, and their model DHO SFs are shown in the bottom panels of Figure~\ref{fig:dho_dist}. 
We also derive an empirical ensemble SF for each point using quasars with a DHO fit falling within 0.1 dex to the selected point, and each ensemble SF is constructed from $\sim5$ quasars.  
The ensemble SFs are computed using Equation 1 of \cite{caplar2017}, and we do not subtract any additional systematic variance as they have been accounted for in our calibration process described in Section~\ref{subsec:err-calib}.
The dashed red line in each bottom panel shows the SF of a comparison best-fit DRW model; the comparison DRW model is obtained by fitting a DRW to light curves used in constructing the corresponding empirical ensemble SF, and taking the median MAP DRW parameters as the best fit. 
The DHO SFs of all three points follow more closely the empirical SFs than the comparison DRW SFs, above the noise floors (grey region).
The noise floor is defined as $\sqrt{2}\,\sigma_{\rm md}$, where $\sigma_{\rm md}$ is the median error of the photometry in the light curve.
The purple contours in the top panels show the distribution of DHO fits in the final overdamped sample of \citetalias{yu2022b}, which matches well with the distribution of DHOs having a SF slope flatter than 0.5.

\subsection{PSDs of Best-fit DHOs}\label{subsec:flat_vs_steep}

\begin{figure*}
    \centering
    \includegraphics[width=0.45\linewidth]{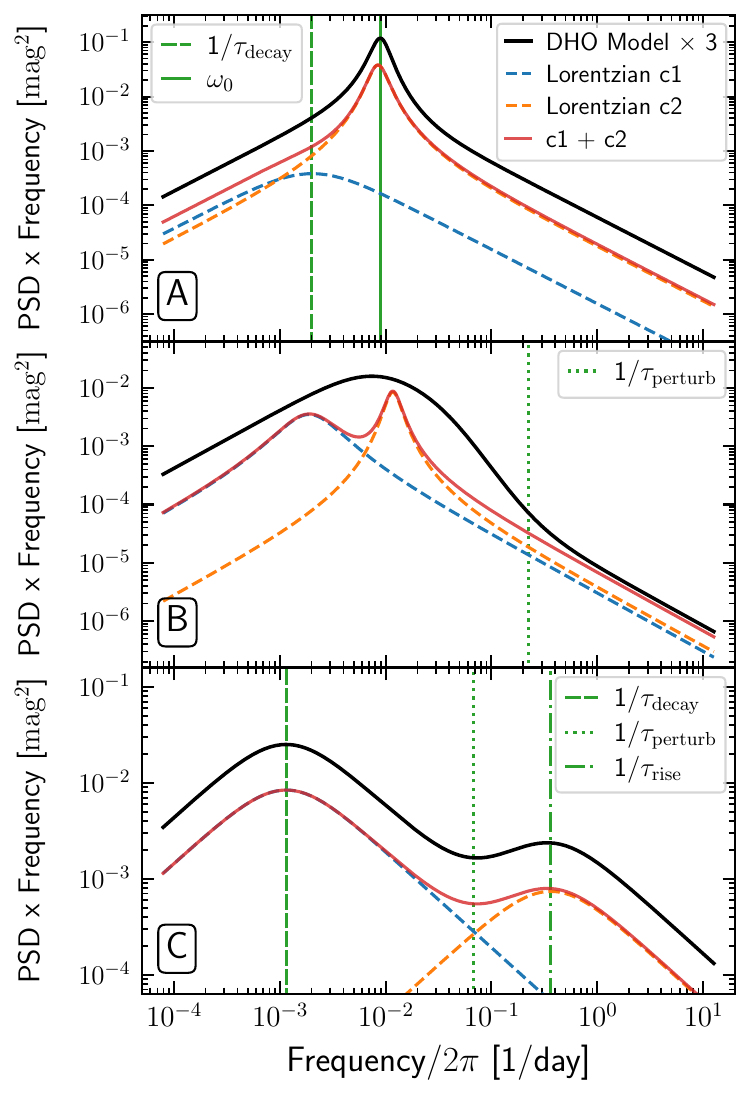}
    \includegraphics[width=0.451\linewidth]{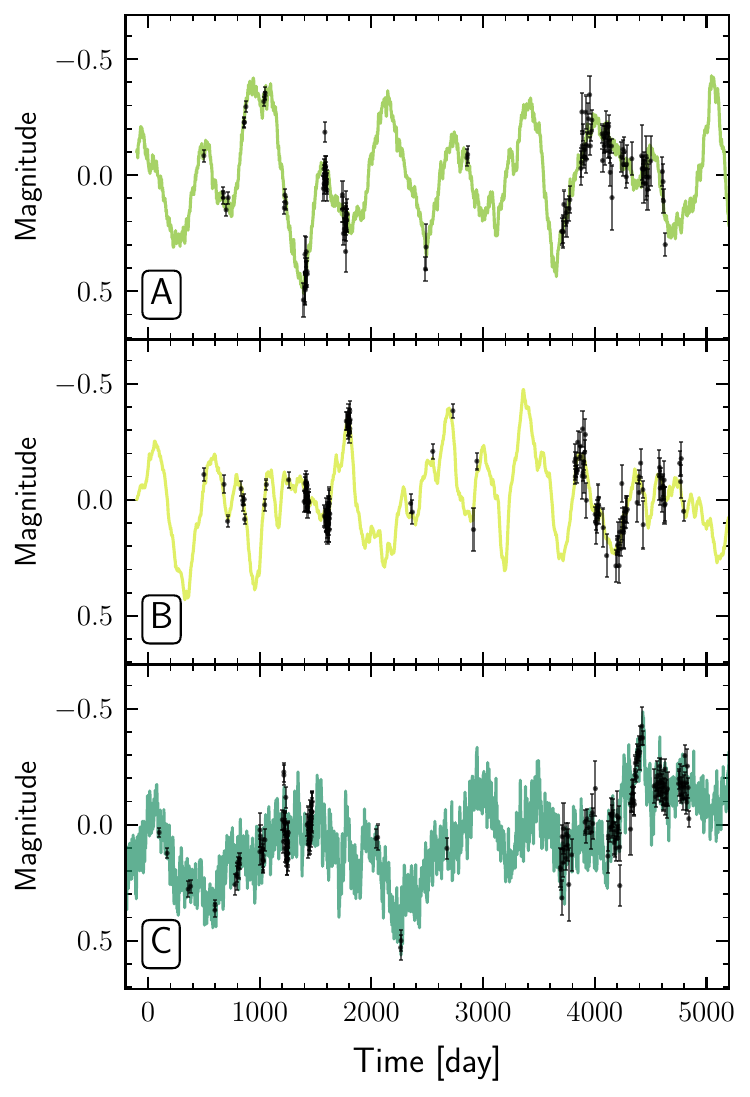}
    \caption{DHO PSDs (left column) and their corresponding simulated light curves (right column) of the three representative points \textbf{A}, \textbf{B}, and \textbf{C} shown in Figure~\ref{fig:dho_dist}.
    {\it Left:} In each panel, the black solid curves show the model DHO PSDs multiplied by 3, the blue and orange dashed curves show the best-fit Lorentzian components, and the red solid curves show the sum of the two Lorentzian components.
    The PSD of \textbf{A} can be well fitted by a sum of one narrow and one broad Lorentzian components. The peak frequency of the narrow component matches the DHO's natural oscillation frequency ($\omega_0$), and peak frequency of the broad component coincides with 1/\taudecay.
    The PSD of \textbf{B} is not well-fitted by the two-Lorentzian model, while 1/\tauperturb~coincides with the frequency where the PSD changes slope from low to high frequencies.
    The PSD of \textbf{C} can be well fitted by a sum of two broad Lorentzian components. 
    The low frequency component peaks at 1/\taudecay, the high frequency component peaks at 1/\taurise, and the dip between the two peaks coincides with 1/\tauperturb.
    {\it Right:} The simulated light curve of \textbf{A} shows quasi-periodic variability, and appears smooth on short timescales. The simulated light curve of \textbf{B} also appears smooth on short timescales, however, it lacks a regular quasi-periodic pattern in comparison to the light curve of \textbf{A}. The simulated light curve of \textbf{C} exhibits excess variability at short timescales in comparison to the light curves of \textbf{A} and \textbf{B}.
    \edit{The black points overplotted on the simulated light curves represent observed light curves of selected \texttt{clean sample} quasars whose best-fit DHO parameters closely match those of \textbf{A}, \textbf{B}, and \textbf{C}, all of which also have similar values of the long-term variability amplitude (\sigmadho).}
    }
    \label{fig:psds}
\end{figure*}

We now explore the connections between DHO parameters and features in DHO model PSDs, using the three representative points \textbf{A}, \textbf{B}, and \textbf{C} shown in Figure~\ref{fig:dho_dist}.
The mode DHO PSD is computed following Equation~\ref{eqn:dho_psd}.
The solid black lines in the left column of Figure~\ref{fig:psds} show the PSDs of \textbf{A}, \textbf{B}, and \textbf{C}, where on the y-axis we plot the product of PSD and frequency (PSD~$\times$~frequency) instead of PSD.
Plotting PSD~$\times$~frequency on the y-axis gives the distribution of the integrated power as a function of frequency~\citep{done2010}, and it also facilitates direct comparisons of UV/optical PSDs to X-ray PSDs~\citep[e.g.,][]{Arevalo2006, McHardy2007, kelly2011}. 
Following the approach often taken in analyzing the PSDs of X-ray light curves by accreting black holes, we fit the model DHO PSD with a sum of two Lorentzian components~\citep[e.g.,][]{Axelsson2005, McHardy2007, done2007}.
The definition of the two-Lorentzian model is adopted from \cite{McHardy2007} and is also shown below,
\begin{equation}
    P(f)=\frac{A_1 Q_{\mathrm{1}} f_{\rm{R}, 1}}{f_{\rm{R},1}^2+4 Q_{1}^2\left(f-f_{\rm{R},1}\right)^2}+\frac{A_2 Q_{2} f_{\rm{R},2}}{f_{\rm{R},2}^2+4 Q_{2}^2\left(f-f_{\rm{R},2}\right)^2}.
\end{equation}\label{eqn:ltz}
\edit{Each Lorentzian component is parameterized by $f_{\rm R}$, $Q$, and $A$, where $f_{\rm R}$ is the resonant frequency, $Q$ is the quality factor defined as $Q = f_{\rm R}/\Delta f_{\rm FWHM}$ with $\Delta f_{\rm FWHM}$ quantifying the full width at half maximum of the Lorentzian, and $A$ is the normalization of the component.}
The peak frequency ($f_{\rm peak}$) of the Lorentzian is related to the resonant frequency by $f_{\rm peak} = f_{\rm R}(1+1/4 Q^2)^{1/2}$.  Note that $Q$ controls the sharpness of the Lorentzian profile in the log-log plot (e.g., Figure~\ref{fig:psds}). The sharper the Lorentzian profile around $f_{\rm peak}$, the larger the $Q$ value.

The PSD of \textbf{A} (top panel in the left column of Figure~\ref{fig:psds}) is well fitted by the sum of a narrow Lorentzian component (orange dashed curve) and a broad Lorentzian component (blue dashed curve).
The peak frequency ($f_{\rm peak}$) of the narrow component is coincident with the derived natural oscillation frequency $\omega_0$ (solid green vertical line) of the DHO. 
For the broad component, we fix its peak frequency to 1/\taudecay~during fitting. We find that allowing the peak frequency of the broad component to freely vary only returns a marginally better fit and the best-fit peak frequency is also very close to 1/\taudecay. 
The top panel in the right column of Figure~\ref{fig:psds} shows a simulated light curve of \textbf{A}, which exhibits regular quasi-periodic variability and appears smooth on short timescales.  
The population represented by point \textbf{A} corresponds the typical underdamped DHOs shown in \citetalias{yu2022b}.
We refer to DHOs represented by point \textbf{A} as `class-A' DHOs (see also DHOs with $\xi < 0.5$ in Figure 14 of~\citetalias{moreno2019}).

The PSD of \textbf{B} (\edit{middle} panel in the left column of Figure~\ref{fig:psds}) looks similar to the PSD of \textbf{A}, but with a much broader peak. 
The center of this broad peak does not coincide with any derived DHO timescales/frequencies, and the two-Lorentzian model cannot generate a good fit to this PSD.
However, 1/\tauperturb~appears to coincide with the frequency where the PSD changes slope from low to high frequencies.
More Lorentzian components might be needed to fit the PSDs of DHOs like \textbf{B}, and we will investigate it further in future work.
The \edit{middle} panel in the right column of Figure~\ref{fig:psds} shows a simulated light curve of \textbf{B}, which also appears smooth on short timescales, but lacks a regular quasi-periodic pattern in comparison to the light curve of \textbf{A}.
The population represented by point \textbf{B} sits between the region typically occupied by underdamped DHOs and the region occupied by overdamped DHOs in the $\xi$--$\omega_0$ parameter space, and this DHO population was not discovered by \citetalias{yu2022b}.
We refer to DHOs represented by point \textbf{B} as `class-B' DHOs.

The PSD of \textbf{C} (bottom panel in the left column of Figure~\ref{fig:psds}) is well fitted by a sum of two broad Lorentzian components.
\edit{During the fitting process, all parameters in the two-Lorentzian model are allowed to vary freely.}
The low frequency broad component (blue dashed curve) peaks at 1/\taudecay, and the high frequency Lorentzian component (orange dashed curve) peaks at 1/\taurise~(see definition in Appendix~\ref{appendix:dho_feats}).
In \citetalias{yu2022b}, \taurise~was interpreted as the timescale taken for an overdamped DHO to respond to a delta-function perturbation injected into the RHS of Equation~\ref{eqn:dho2}. 
The dip between the two peaks coincides with 1/\tauperturb, which is the frequency when the power from the high frequency Lorentzian component exceeds that from the low frequency component (the orange curve is on top of the blue curve), as the frequency increases. 
The bottom panel in the right column of Figure~\ref{fig:psds} shows a simulated light curve of \textbf{C}, which exhibits excess variability on short timescales in comparison to the light curves of \textbf{A} and \textbf{B}.
The population represented by point \textbf{C} corresponds the typical overdamped DHOs shown in \citetalias{yu2022b}.
We refer to DHOs represented by point \textbf{C} as `class-C' DHOs.

\subsection{Class-C DHO Parameters vs. \lambdarf, \lthreezero, and \fhmgii}\label{subsec:phys_corr}

In this section, we present the correlations between best-fit DHO parameters and quasar physical properties, specifically \lambdarf, \lthreezero, and \fhmgii.
\edit{These correlations are derived using the \texttt{core sample}, within which 957 $g$-band fits and 957 $r$-band fits are classified as class-C DHOs.}
\lambdarf~serves as a crude proxy for the distance to the central SMBH from where the UV/optical emission originates in the accretion disk~\citep{shakura1973}
\lthreezero~probes the quasar's bolometric luminosity (\lbol) assuming a fixed bolometric correction~\citep[e.g.,][]{richards2006}. 
\lthreezero~and $\rm FWHM_{Mg II}$ together probes the mass of the SMBH~\citep[e.g.,][]{shen2011, Shen2024}.
Since we do not have a sufficiently large number of class-A and class-B DHOs (\edit{totaling $\sim25\%$} of the \texttt{core sample}), we only focus on investigating the correlations between quasar physical properties and class-C DHO parameters.

\begin{figure}
    \centering
    \includegraphics[width=0.98\linewidth]{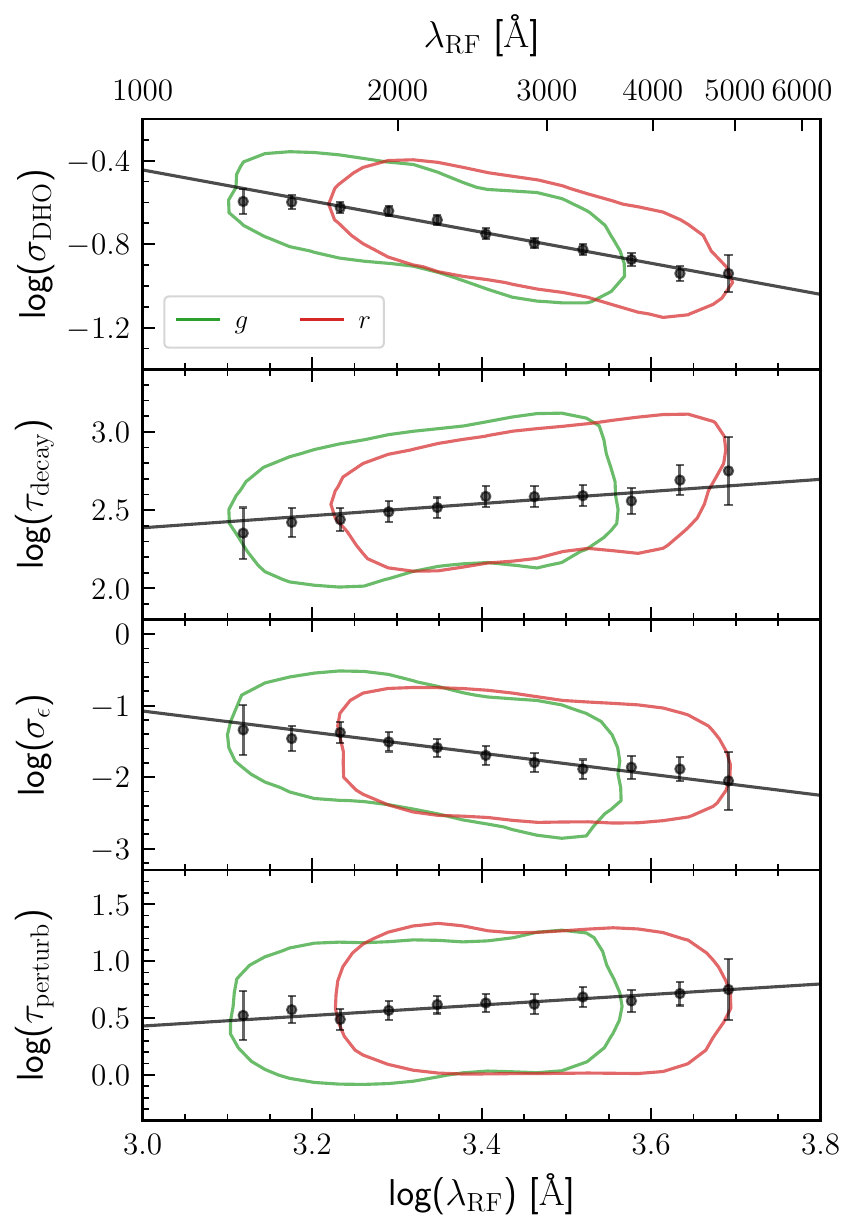}
    \caption{The wavelength dependence of best-fit parameters of class-C DHOs. 
    The best-fit DHO parameters (y-axis) have been redshifted to the rest-frame of their quasars, as well as normalized to log(\lthreezero) = 45 and log(\fhmgii) = 3.6 using the coefficients reported in Table~\ref{tab:dhopropfit}. 
    The green and red contours in each panel give the distribution of the $g$-band and $r$-band best-fit DHO parameters in the \texttt{core sample}, respectively.
    \sigmadho, \tauperturb, \sigmanoise, and \tauperturb~exhibit nearly monotonic correlations with \lambdarf. 
    The black points in each panel show the binned average with their error bars estimated through bootstrapping. 
    }
    \label{fig:lambda_dep}
\end{figure}

\begin{figure*}
    \includegraphics[width=1\linewidth]{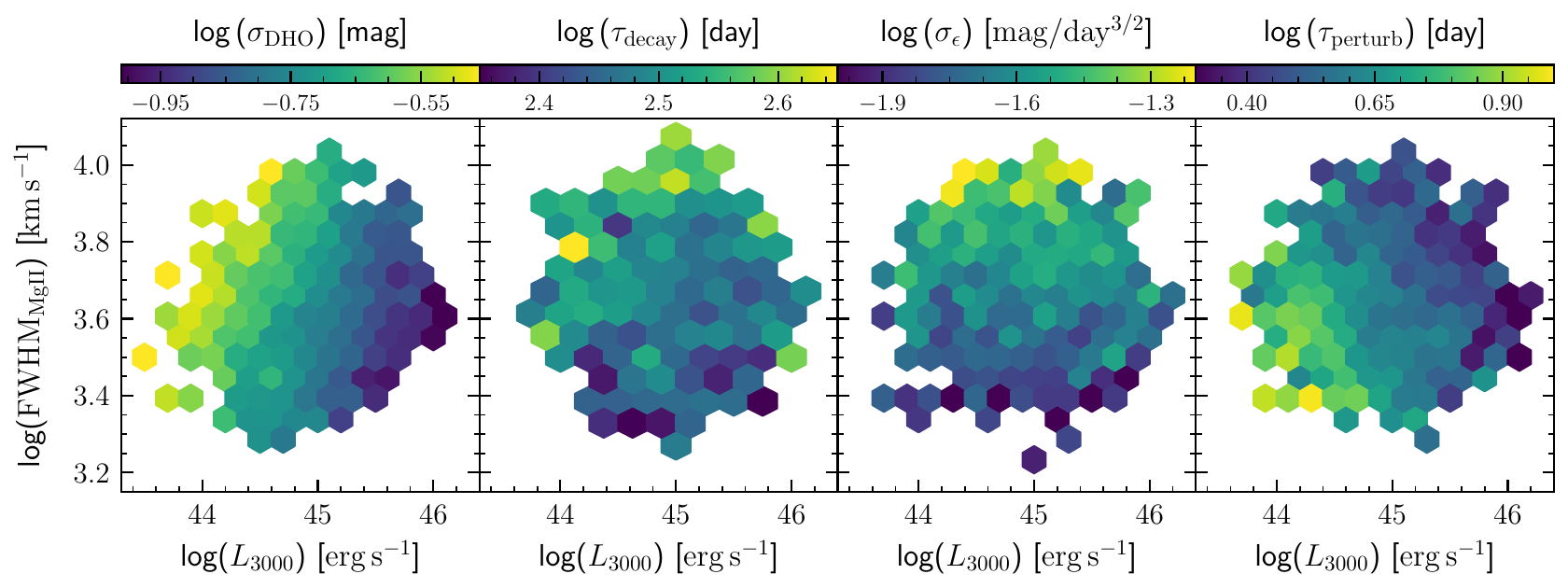}
    \caption{Trends of \edit{\texttt{clean sample}} DHO parameters with \lthreezero~and \fhmgii. 
    We have redshifted the best-fit DHO parameters to the rest-frame of their quasars as well as normalized them to log(\lambdarf)~$=3.4\,(\sim2500$\AA)~using the coefficients listed in Table~\ref{tab:dhopropfit}.
    \sigmadho, \sigmanoise, and \tauperturb, show clear correlations with both \lthreezero~and \fhmgii.
    \taudecay~shows a weak correlation with \fhmgii~and no correlation with \lthreezero.
    }
    \label{fig:hexbin}
\end{figure*}

We perform bi-sector regressions to determine the correlations between DHO parameters and quasar physical properties using \texttt{core sample} DHO fits.
We perform the regression following the equation, 
\begin{eqnarray}\label{eqn:propCorr}
    \mathrm{log}(X) &&= A\,{\rm log(\lambda_{RF})} + B\,{{\rm log}(L_{3000})}\nonumber\\
                    &&+\, C\,{\rm log(FWHM_{Mg\,II})} + D,
\end{eqnarray}
where $X$ represents one of \sigmadho, \taudecay, \sigmanoise, and \tauperturb.
The best-fit coefficients and their statistical uncertainties are listed in Table~\ref{tab:dhopropfit}. 

\begin{deluxetable*}{ccccc}
    \tablecaption{Best-fit coefficients and statistical uncertainties for Equation~\ref{eqn:propCorr}}
    \tablehead{\colhead{$X$}   &   \colhead{\hspace{1cm}$A\,$(\lambdarf)}\hspace{1cm} & \colhead{\hspace{1cm}$B\,$(\lthreezero)}\hspace{1cm}  & \colhead{\hspace{1cm}$C\,$(\fhmgii)}\hspace{1cm} & \colhead{\hspace{0.2cm}$D$}\hspace{1.cm}}
    \startdata
    \sigmadho       &    -0.746~$\pm$~0.030  &  -0.242~$\pm$~0.007 & ~0.362~$\pm$~0.018 & 11.348~$\pm$~0.404\\
    \taudecay       &    ~0.388~$\pm$~0.083  &  -0.025~$\pm$~0.019 & ~0.169~$\pm$~0.050 & ~1.679~$\pm$~1.114\\
    \sigmanoise     &    -1.444~$\pm$~0.146  &  ~0.007~$\pm$~0.032 & ~0.897~$\pm$~0.093 & -0.368~$\pm$~1.892\\
    \tauperturb     &    ~0.448~$\pm$~0.092  &  -0.190~$\pm$~0.021 & -0.385~$\pm$~0.060 & 
    ~9.008~$\pm$~1.220\\
    \enddata
\end{deluxetable*}\label{tab:dhopropfit}

DHO parameters, \sigmadho, \taudecay, \sigmanoise, and \tauperturb, exhibits correlations with \lambdarf.
Figure~\ref{fig:lambda_dep} shows the correlations of \sigmadho, \taudecay, \sigmanoise, and \tauperturb~with \lambdarf.
\sigmadho~and \sigmanoise~decrease with increasing \lambdarf, while \taudecay~and \tauperturb~increase with increasing \lambdarf.
The best-fit DHO parameters are normalized to log(\lthreezero) = 45 and log(\fhmgii) = 3.6 using the coefficients listed in Table~\ref{tab:dhopropfit}.
The black lines in each panel of Figure~\ref{fig:lambda_dep} show the best-fit relation of \sigmadho, \taudecay, \sigmanoise, and \tauperturb~with \lambdarf.
The green and red contours show the distribution of \texttt{core sample} $g$-band and $r$-band DHO parameters, respectively. 
The correlations of DHO's long-term variability amplitude (\sigmadho) and characteristic timescale (\taudecay) with \lambdarf~are similar to the correlations of DRW's asymptotic amplitude and damping timescale with \lambdarf, as discovered by other investigations~\citep[e.g.,][]{macleod2010, suberlak2021, stone2022}. 

DHO parameters \sigmadho, \taudecay, \sigmanoise, and \tauperturb~also display correlations with \lthreezero~and \fhmgii.
Figure~\ref{fig:hexbin} presents these correlations using \texttt{clean sample} DHO fits, where the DHO parameters have been normalized to log(\lambdarf)~$=3.4\,(\sim2500$\AA)~using the coefficients listed in Table~\ref{tab:dhopropfit}.
The color scale indicates the median DHO parameters for quasars that fall into each hex bin.
\sigmadho, \sigmanoise, and \tauperturb, show clear correlations with both \lthreezero~and \fhmgii. The correlations of \sigmadho~with \lthreezero~and \fhmgii~is the strongest, possibly because \sigmadho~is best constrained with a median uncertainty of 0.14 for log(\sigmadho). 
The correlations of \sigmanoise~and \tauperturb~with \lthreezero~and \fhmgii~are also strong, but their trends are not as systematic---this may owe to their relatively large uncertainties, which are 0.62 and 0.4 for log(\sigmanoise)~and log(\tauperturb), respectively.
\taudecay~shows a weak correlation with \fhmgii~and no correlation with \lthreezero.

\begin{table}
    \centering
    \caption{Best-fit coefficients and statistical uncertainties for Equation~\ref{eqn:propCorr2}}
    \vspace{-.3cm}
    \begin{tabular}{cccc}
    \hline\hline
    $X$   &   $E\,$(\lledd) & $F\,$(\mbh) & $G$\\
    \hline
    \sigmadho       &    -0.382~$\pm$~0.010  &  -0.270~$\pm$~0.008 & ~3.705~$\pm$~0.066\\
    \taudecay       &    -0.087~$\pm$~0.019  &  -0.031~$\pm$~0.016 & ~1.305~$\pm$~0.129\\
    \sigmanoise     &    -0.336~$\pm$~0.035  &  ~0.025~$\pm$~0.028 & ~2.534~$\pm$~0.230\\
    \tauperturb     &    -0.376~$\pm$~0.024  &  -0.405~$\pm$~0.020 & ~2.180~$\pm$~0.164\\
    \hline
    \end{tabular}
    \label{tab:dhoprop2}
\end{table}

We also explore the correlations of DHO parameters with \lledd~and \mbh.
We directly fit DHO parameters as a function of \lledd~and \mbh,
\begin{eqnarray}\label{eqn:propCorr2}
    \mathrm{log}(X) &&= A\,{\rm log(\lambda_{RF})} + E\,{{\rm log}(L/L_{\rm Edd})} \nonumber\\
                    &&+ F\,{{\rm log}(M_{\rm BH})} + G,
\end{eqnarray}
where the value of A is directly taken from Table~\ref{tab:dhopropfit}.
We assume a bolometric correction of 5.15~\citep{richards2006} for \lthreezero, and adopt the recipe for \mbh~estimation from \cite{Shen2024}. The best-fit coefficients are listed in Table~\ref{tab:dhoprop2}.
The anti-correlation of \sigmadho~with \lledd~and \mbh~is consistent with the result of \citetalias{yu2022b}, and similar to the conclusions of others who used different methods to quantify quasar long-term variability amplitude~\citep[e.g.,][]{wilhite2007, bauer2009, macleod2010, simm2016, sanchez-saez2018, suberlak2021, kang2021, tang2023, arevalo2023, Petrecca2024}. 

\subsection{Comparison to Y22}\label{subsec:vs_yu22}

The distribution of best-fit DHO parameters as well as their correlations with quasar physical properties found here are broadly consistent with those found by \citetalias{yu2022b}.
Specifically, we find similar wavelength dependencies of \sigmadho, \sigmanoise~to those reported in \citetalias{yu2022b}, as well as similar correlations of \sigmadho~with \lledd~and \mbh. 

There are also a few differences between the results shown here and those by~\citetalias{yu2022b}, as well as new findings.
Firstly, the correlation of \tauperturb~with \lambdarf~no longer shows a clear break at $2500$\AA, we suspect the break shown in \citetalias{yu2022b} could be a combination of a positive correlation of \tauperturb~with \lambdarf~and an anti-correlation of \tauperturb~with both \lthreezero~and \mbh. 
After normalizing \tauperturb~to a fixed value of \lthreezero~and a fixed value of \mbh, \tauperturb~shows a monotonic dependence on \lambdarf~(see the bottom panel of Figure~\ref{fig:lambda_dep}). 
Secondly, in \citetalias{yu2022b}, \tauperturb~did not show a clear correlation with \mbh~before it was normalized by the \mbh~of its corresponding quasar. In this work, \tauperturb~shows a strong anti-correlation with \mbh~(see Table~\ref{tab:dhoprop2}). 
We suspect this difference may arise from \citetalias{yu2022b} only looking at \tauperturb\ values that fell into a narrow range of \lambdarf~(i.e., $\sim2240$\rmAA~to~$\sim2820$\rmAA), thus limiting the range of \tauperturb~examined.
Lastly, we discover a new population of DHOs (class-B) that sits between the cluster of typical underdamped DHOs (class-A) and the cluster of typical overdamped DHOs (class-C) in the $\xi$--$\omega_0$ parameters space (see Figure 3 of~\citetalias{yu2022b}). 
class-B DHOs, as represented by point \textbf{B} in Figure~\ref{fig:dho_dist} and Figure~\ref{fig:psds}, feature a SF slope steeper than that of a DRW.

\section{Discussion}\label{sec:discussion}

\begin{figure}
    \centering
    \includegraphics[width=\linewidth]{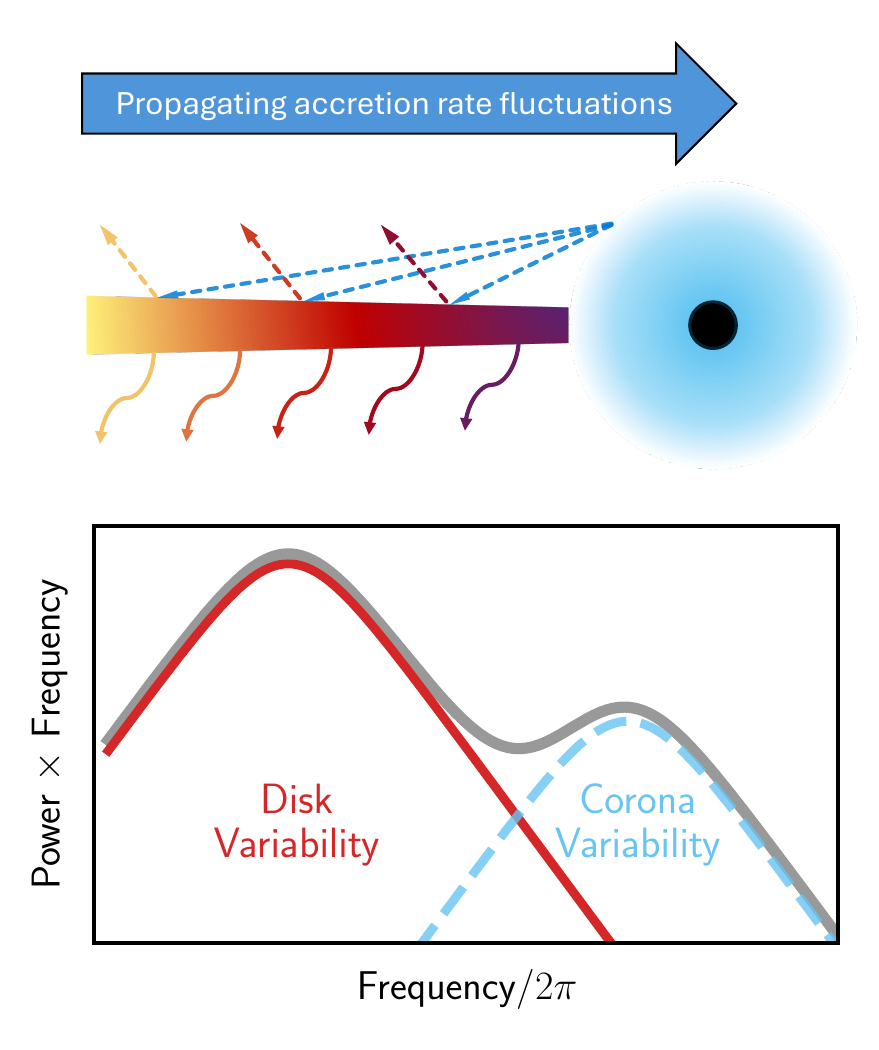}
    \caption{A schematic diagram showing the accretion flow geometry (top) and the suggested contributions of different accretion flow components to the PSD of the quasar UV/optical variability captured by class-C DHOs (bottom).
    {\it Top:} A side cross-section of the accretion disk, the inner hot accretion flow, and the central SMBH, are shown.
    The blue arrow at the top shows the direction of the inwardly-propagating accretion rate fluctuations.
    The solid arrows coming out of the accretion disk represent the intrinsic disk emission.
    The dashed arrows hitting and coming out of the accretion disk represent coronal emission reprocessed by the disk.
    {\it Bottom:} In the PSD, the low-frequency bump (red) owes to intrinsic variability from the accretion disk, while the high-frequency bump (blue) owes to variability from the inner hot flow. In UV/optical wavelengths, one observes both the intrinsic disk variability (solid arrows in the top panel) and the reprocessed X-ray variability (dashed arrows in the top panel) associated with the inner hot flow. 
    }
    \label{fig:flow_diagram}
\end{figure}

\subsection{AGN UV/optical Variability Captured by\\ Class-C DHOs}\label{subsec:discuss_flat}

The PSDs of class-C DHOs can be well-fitted by two broad Lorentzian components, which suggests that the variability associated with each Lorentzian component might have different physical origins.
PSDs featuring two or multiple Lorentzian components have been observed in the X-ray light curves of accreting stellar-mass black holes~\citep[e.g.,][]{Nowak2000, Churazov2001, Axelsson2005, done2007} as well as of a limited number of AGN~\citep[e.g.,][]{McHardy2007, Alston2019}.
In particular, \cite{McHardy2007} measured the frequency-resolved lags between the soft ($0.5$--$2$ keV) and hard ($2$--$8.8$ keV) band X-ray light curves of Ark 564, the PSDs of which are also well-fitted by two broad Lorentzian components. 
They found that the high-frequency (HF) component of the hard band light curve has no lag from its soft band counterpart, while the low-frequency (LF) component of the hard band light curve has a positive lag behind its soft band counterpart. 
They concluded that the variability associated with the HF and LF components might originate from different regions in the accretion flows.
Similar to this picture, we suggest that the LF component (peaking at 1/\taudecay) of class-C DHOs probes the intrinsic variability of the accretion disk, which might be driven by propagating accretion rate fluctuations~\citep{Lyubarskii1997} or local thermal fluctuations~\citep[e.g.,][]{dexter2011a, ruan2014, sun2020}, and the HF component (peaking at 1/\taurise) probes the reprocessed X-ray variability originating from the hot accretion flow near the SMBH.

Indeed, AGN UV/optical PSDs featuring two broad components were recently predicted by a physical model developed from first principles~\citep{Hagen2024}.
\cite{Hagen2024} assume that the variability intrinsic to the disk follows the propagating fluctuations model~\citep{Lyubarskii1997}, which results in a broad Lorentzian component in the PSD with a peak frequency of $\nu_{\rm gen}$. $\nu_{\rm gen}$ is a characteristic frequency of the local accretion rate fluctuation. 
The fluctuations at each radius then propagate inward and grow in amplitude following a multiplicative manner~\citep{uttley2005, Arevalo2006}.
\cite{Hagen2024} also assume that the intrinsic variability of the inner hot flow (i.e., X-ray corona) is driven by these propagating accretion rate fluctuations, but the Lorentzian corresponding to the reprocessed X-ray-variability peaks at a much higher frequency due to the compact size of the X-ray corona~\citep{Ingram2011}.
Therefore, the observed UV/optical variability consists of two components, one representing the variability intrinsic to the disk, and another representing the reprocessed variability of the X-ray corona.
Figure~\ref{fig:flow_diagram} displays a schematic diagram of this picture.
The top panel of Figure~\ref{fig:flow_diagram} shows the accretion flow geometry, and the bottom panel shows the suggested contributions of different accretion flow components to the PSD of quasar UV/optical variability modeled by class-C DHOs.
The propagating fluctuations model is also consistent with the recent findings of~\cite{neustadt2022}, who suggested that the temperature fluctuations of the accretion disk is dominated by slow in-going waves.


\subsection{Trends of Class-C DHO Parameters with \lambdarf~and AGN Physical Properties}\label{subsec:discuss_long_param_trends}

\begin{figure}
    \centering
    \includegraphics[width=1\linewidth]{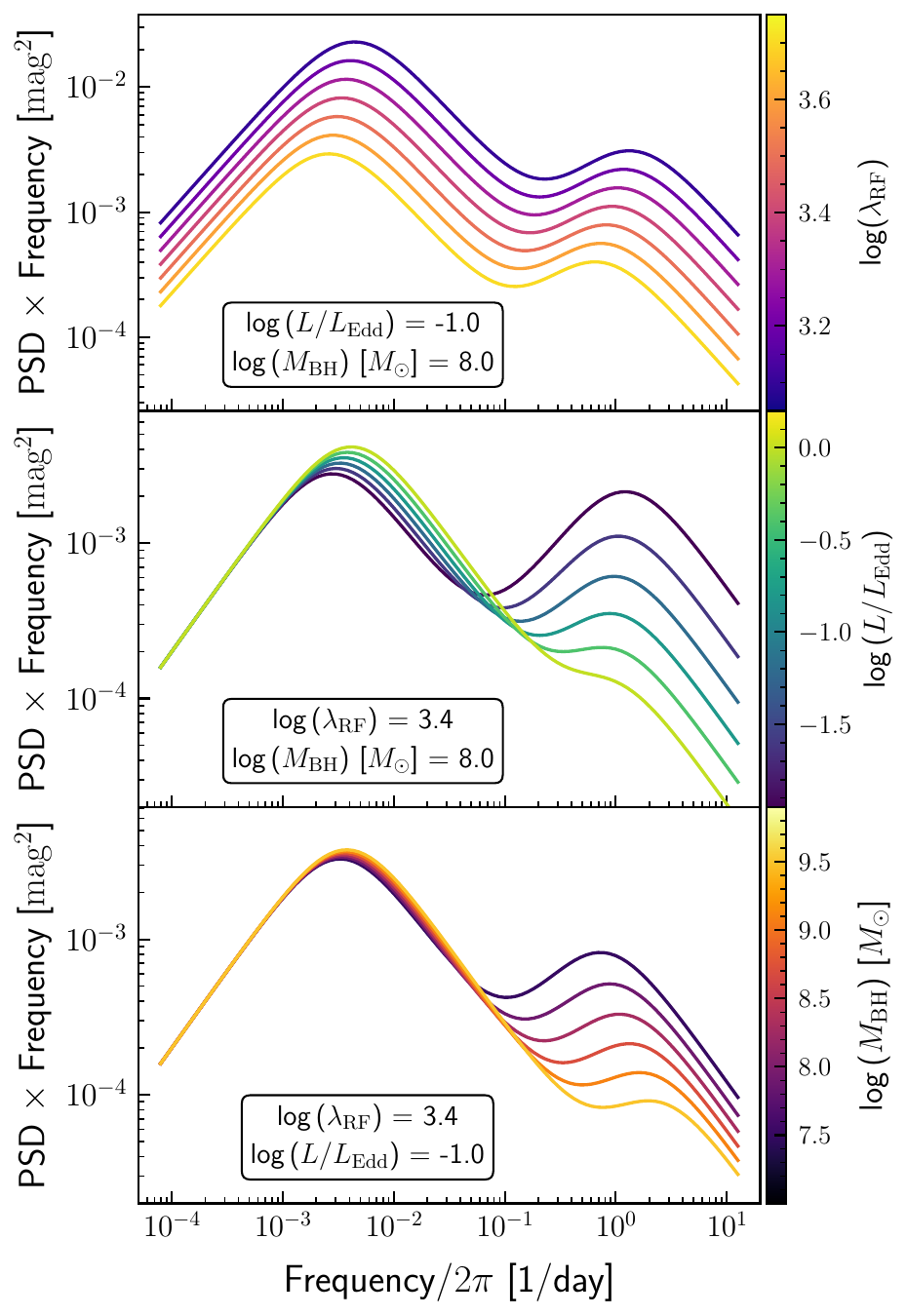}
    \caption{Simulated class-C DHO PSDs as a function of \lambdarf, \lledd, and \mbh.
    \edit{The input DHO parameters are computed using Equation~\ref{eqn:propCorr2}, with coefficients provided in Table~\ref{tab:dhopropfit} and Table~\ref{tab:dhoprop2}.}
    {\it Top}: \edit{DHO PSDs at log(\lambdarf) $=3.1$--$3.7$ in steps of 0.1 dex, given log(\lledd)~$= -1.0$ and log(\mbh) $= 8.0$.}  
    The variability power decreases as \lambdarf~increases, at all frequencies shown.
    {\it Middle}: \edit{DHO PSDs at log(\lledd)~$=-2.0$--$0.0$ in steps of 0.4 dex, given log(\lambdarf)~$=3.4$ and log(\mbh)~$=8$, and each PSD is normalized relative to its power at a frequency $\simeq0$.}
    The relative short-term variability power decreases significantly with an increasing log(\lledd).
    {\it Bottom}: \edit{DHO PSDs at log(\mbh)~$=7.5$--$9.5$ in steps of 0.4 dex, given log(\lambdarf)~$=3.4$ and log(\lledd) $=-1.0$, and each PSD is normalized relative to its power at a frequency $\simeq0$.} Similar to the middle panel, the relative short-term variability power decreases with log(\mbh). 
    The peak frequency of the HF component increases with log(\mbh), but we note that our light curve cadence is not sufficient to robustly constrain this variability feature, so the trend shown could be an artifact.  
    }
    \label{fig:sim_psd}
\end{figure}

The correlations of class-C DHO parameters with \lambdarf~can be understood from the expectation that shorter wavelength photons are emitted by accretion disk regions that are closer to the SMBH than that of longer wavelength photons.
The top panel of Figure~\ref{fig:sim_psd} shows simulated DHO PSDs at different wavelengths for quasars with a fiducial log(\lledd) $= -1.2$ and log(\mbh) $= 8.8$.
The DHO parameters are computed using Equation~\ref{eqn:propCorr2}, with their coefficients taken from Table~\ref{tab:dhopropfit} and Table~\ref{tab:dhoprop2}.
The most noticeable trend in the top panel of Figure~\ref{fig:sim_psd} is an increasing variability power (y-axis) with \lambdarf, across the full range of frequencies.
In the propagating fluctuations model, accretion rate fluctuations in the accretion disk grow in amplitude (\sigmadho) as they propagate towards the SMBH, resulting in a higher power for the associated LF Lorentzian component.
Similarly, as the distance to the SMBH decreases, the influence of the central X-ray corona increases, resulting in stronger reprocessed variability and a higher power for the associated HF Lorentzian component.
The decreasing peak frequency of the LF Lorentzian component (1/\taudecay) with \lambdarf~can be linked to the increasing characteristic timescales of the local disk with radius~\citep{frank2002}.
The decreasing peak frequency of the HF Lorentzian component with \lambdarf~is more puzzling, as we might expect the variability timescale of the X-ray corona to be independent of the reprocessed photon wavelengths.
However, we note that our light curve cadence ($\sim1$ day) is insufficient to robustly constrain the peak frequency (1/\taurise) of the HF component for most of our quasars, and thus we did not put a hard limit on the inferred \taurise~in Section~\ref{subsec:clean_sample}.
More frequently-sampled light curves are needed to robustly constrain the correlation of the HF-component peak frequency with \lambdarf.

DHO long-term variability amplitude (\sigmadho) shows an anti-correlation with both \lledd~and \mbh~of our quasars.
This anti-correlation is expected in the propagating accretion rate fluctuations picture.
The flux emitted per unit area from the accretion disk as a function of distance ($R$) to the central SMBH is
\begin{eqnarray}\label{eqn:disk_bb}
    F(R) = \sigma T(R)^{4}\,&&\propto\,L/L_{\rm Edd}\,M_{\mathrm{BH}}^{2}R^{-3},
\end{eqnarray}
where $T(R)$ is the effective disk temperature at $R$~\citep{frank2002}.
From Equation~\ref{eqn:disk_bb}, the radius of the disk annulus emitting with a fixed blackbody temperature $T(R)$ will increase with increasing \lledd~and \mbh.
Since the variability amplitude decreases as $R$ increases (moving away from the SMBH),~\sigmadho~should scale negatively with  \lledd~and \mbh. 
We note that pure X-ray reprocessing can also explain the anti-correlation of variability amplitude with \lledd, where a higher \lledd~leads to a lower X-ray to UV flux ratio given a fixed~\mbh \citep[e.g.,][]{Vasudevan2007, kubota2018, ruan2019}, and thus a lower relative amplitude for the reprocessed variability.

\taudecay~displays a weak anti-correlation with \lledd, and nearly no correlation with \mbh.
If the long-term variability of quasars is driven by accretion rate fluctuations, one would expect its characteristic timescale \taudecay~to correlate with the local viscous timescale ($t_{\rm visc}$).
However, $t_{\rm visc}$ is predicted to scale with \lledd, \mbh, and \lambdarf~as $(L/L_{\rm Edd})^{1/2}\,M_{\rm BH}^{1/2}\,\lambda_{\rm RF}^2$~\citep{shakura1973}.
One possible explanation for the discrepancy is that, in reality, a range of disk radii covering a range of intrinsic timescales contributes to the observed flux and variability of each photometric band, which leads to a smoothing effect---a smaller exponent for the power-law correlation with \lambdarf~\citep[e.g.,][]{macleod2010, zhou2024}.
Indeed, once we fix the scaling of \taudecay~with \lambdarf~to a power of 2 (as opposed to the best-fit value of $0.388$), we get a correlation between \taudecay~and \lledd~and \mbh~to a power of $\sim0.35$ and $\sim0.4$, respectively---closer to the predicted exponent of $0.5$.
We also note that the expected $t_{\rm visc}$ for our quasars from the standard thin disk model are much longer than their best-fit \taudecay.
Recent simulations have shown that the viscous/inflow timescale can be much shorter than that expected for a standard thin disk if the disk is dominated by magnetic pressure and the disk scale-height is non-negligible~($>$ 0.1)~\citep[][]{Hopkins2024a, Hopkins2024b}. 
In addition, there are recent observational evidence supporting thicker AGN accretion disks~\citep[e.g.,][]{yao2022}.


The anti-correlations of \sigmanoise~and~\tauperturb~with \lledd~and \mbh~may be probing a decrease in X-ray reprocessing with \lledd~and \mbh.
The middle panel of Figure~\ref{fig:sim_psd} shows simulated DHO PSDs for a range of log(\lledd), at a fiducial log(\lambdarf)~$=3.4$~($\sim2500$\AA) and log(\mbh)~$=8$.
The DHO parameters are computed using Equation~\ref{eqn:propCorr2} with their coefficients taken from Table~\ref{tab:dhopropfit} and Table~\ref{tab:dhoprop2}.
The PSDs are normalized to their values at $f\simeq0$, which helps examine the change in the short-term variability as function of \lledd.
The power of the short-term variability (HF bumps) decreases significantly as \lledd~increases.
This trend can be expected if the accretion flow geometry changes as a function of \lledd; more specifically, a larger \lledd~means that less accretion power is dissipated in the central hot flow, leading to a decrease in X-ray reprocessing~\citep[e.g.,][]{kubota2018, giustini2019, Hagen2024b}. 
The bottom panel of Figure~\ref{fig:sim_psd} shows simulated DHO PSDs at a range log(\mbh) given log(\lambdarf)~$=3.4$~($\sim2500$\AA) and log(\lledd)~$=-1$.
Similar to the trend in the middle panel, the short-term variability power decreases as the \mbh~increase, which suggests that X-ray reprocessing is more prominent in AGN with smaller \mbh. 
This trend is consistent with the observed anti-correlation between X-ray variability amplitude and \mbh, if one assumes that the short-term variability (HF Lorentzian component) of class-C DHOs is mostly driven by X-ray reprocessing~\citep[e.g.,][]{Lu2001, O'Neill2005, Kelly2013}
We note that the positive correlation of HF-component peak frequency (1/\taurise) with \mbh~might be an artifact, because our light curves are not dense enough to robustly constrain the associated DHO timescale (\taurise).
Finally, the general trend of increasing short-term variability power with \lledd~and~\mbh~is also consistent with the results of a recent work by~\cite{Arevalo2024}, who used ensemble PSDs to reveal similar trends~(see Figures 2 and 3 of~\citealt{Arevalo2024}).

\subsection{AGN UV/optical Variability Captured by\\ Class-A and Class-B DHOs}\label{subsec:discuss_steep}
Class-A and class-B DHOs describe intrinsic variability that is characterized by suppressed power at short timescales in comparison to the DRW model.
The PSDs of class-A DHOs (represented by point \textbf{A} in Figure~\ref{fig:psds}) show strong QPO-like signatures, while the PSDs of class-B DHOs (represented by point \textbf{B} in Figure~\ref{fig:psds}) lack QPO-like signatures.
The simulated light curves of both types are smoother on short timescales than of that class-C DHOs~(right column of Figure~\ref{fig:psds}). 

The PSDs of class-A DHOs are well fitted by a sum of one broad and one narrow Lorentzian components.
The combination of one broad and one narrow Lorentzian components is often used to model the PSDs of X-ray binaries that exhibit QPO signatures~\citep[e.g.,][]{Ingram2011, Ingram2012}.
\edit{For the PSDs of X-ray binaries, the broad component is typically associated with the variability driven by the propagation of fluctuations in the accretion flow, and the narrow component is often believed to arise from Lense-Thirring precession of the inner hot accretion flow (i.e., corona)~\citep{ingram2009, Ingram2011, Ingram2013}.}
In a similar fashion, we suggest that the broad component in the PSDs of class-A DHOs could be associated with the long-term variability intrinsic to accretion disk, while the narrow component could be associated with X-ray reprocessing of the inner hot flow that undergoes Lense-Thirring precession~\citep{Veledina2015a, Veledina2015b}. 
Alternatively, this QPO-like signature could arise from two SMBH orbiting each other in a binary~\citep{begelman1980}.

The PSDs of class-B DHOs cannot be fitted by a two-Lorentzian model.
By comparing its simulated light curve with that of class-A and class-C DHOs (right column of Figure~\ref{fig:psds}), we suspect that this population of DHOs might have several origins.
For example, they could be intrinsically class-A DHOs, but the big gaps in our light curves prevent the DHO model from capturing the QPO signal, while still capturing the smooth variability at short timescales. 
Alternatively, they may be associated with quasars undergoing extreme accretion episodes (e.g., changing-state quasars), where excess variability at timescales of decades leads to steeper SFs~\citep[e.g.,][]{Ruan2016, Noda2018, Graham2020}.

\section{Summary \& Conclusion}\label{sec:conclude}
We modeled the UV/optical variability of $21,767$ SDSS Stripe 82 quasars as a noise-driven damped harmonic oscillator (DHO) process, and applied it to 22-year-long light curves collected from SDSS, PS1, and ZTF. 
For $2,888$ quasars with clean $g$-band DHO fits and $1,861$ quasars with clean $r$-band DHO fits, we examined the variability properties captured by our DHO model using its SF and PSD.
We searched for the correlations of four DHO parameters---\sigmadho, \taudecay, \sigmanoise, and \taudecay~with \lambdarf, \lthreezero~and \fhmgii~of the observed quasars.  
Our results are summarized as follows:

\begin{itemize}
    \item A DHO process can capture a diverse spectrum of quasar UV/optical variability revealed by the  empirical SF of the light curves. The majority of the quasar light curves that are well-fit by a DHO had SFs with a slope flatter than that of a DRW (i.e., 0.5). 
    \edit{About $25\%$} of our best-fit DHOs had SFs with a slope steeper than 0.5, among which some exhibit QPO-like signatures.
    
    \item The PSDs of best-fit DHOs displaying flat SFs (class-C) can be well-modeled by a sum of two broad Lorentzian components. The low frequency Lorentzian component peaks at 1/\taudecay, and the high frequency component Lorentzian peaks at 1/\taurise~(see Figure~\ref{fig:psds}). 
    
    \item The PSDs of best-fit DHOs exhibiting QPO-like signatures (class-A) can be well modeled by a sum of one narrow and one broad Lorentzian components. The peak frequency of the narrow Lorentzian component coincides with DHO's natural oscillation frequency ($\omega_0$), and the peak frequency of the broad component coincides with 1/\taudecay~(see Figure~\ref{fig:psds}).
    
    \item All four DHO parameters of class-C DHOs---\sigmadho, \taudecay, \sigmanoise, and \tauperturb---exhibit correlations with \lambdarf, \lthreezero~and \fhmgii~(and/or \lledd~and \mbh). In particular, \sigmadho~anti-correlates strongly with both \lledd~and \mbh, \taudecay~anti-correlates weakly with \lledd, \sigmanoise~shows a strong anti-correlation with \lledd, and \sigmadho~shows a strong anti-correlation with both \lledd~and \mbh. 

    \item We suggest that quasar UV/optical variability well-modeled by the class-C DHO can be explained using a scenario in which propagating accretion rate fluctuations in the disk give rise to the observed long-term variability (i.e., the LF Lorentzian component in the PSD), while reprocessing of X-ray photons from the central corona gives rise to the short-term variability (i.e., the HF Lorentzian component in the PSD).

    \item Given the temporal sampling rate of our light curves, the recovery of the long-term variability parameters, \sigmadho~and \taudecay, are affected by the light curve baseline, whereas the recovery of the short-term variability parameters, \sigmanoise~and \tauperturb~are less affected by the light curve baseline.

\end{itemize}


Results of this work demonstrate that stochastic models can well capture the intrinsic variability properties encoded in quasar light curves. 
Discovered correlations between variability properties extracted via stochastic modeling and quasar fundamental properties could also power new methods that will derive quasar fundamental properties using time-domain data alone. 

The Vera C. Rubin Observatory Legacy Survey of Space and Time~\citep[LSST;][]{ivezic2019} will soon begin monitoring tens of millions of AGN with an average cadence of 3 days for 10 years. 
The analysis presented here can be quickly applied to LSST quasars that also have existing archival light curves from SDSS, PS1, and ZTF, assuming their photometry are well-calibrated to each other.
The superb photometric S/N and temporal cadence of LSST light curves will allow better constrains on the PSD of class-C DHOs at high frequencies, as well as the correlations of class-C DHO PSDs with quasar fundamental properties (e.g., the trends shown in Figure~\ref{fig:sim_psd}).
The finding that the recovery of \sigmanoise~and \tauperturb~does not depend strongly on super-long light curves suggests their potential ability to enable AGN/quasar selections using early LSST data releases~\citep[][]{Savic2023}.

\vspace{.5cm}
\edit{We thank the anonymous referee for the thoughtful comments that helped us improve the paper.}
We thank Neven Caplar, Chris Done, Scott Hagen, and Matthew Temple for useful discussions.
W.Y. acknowledges support from the Dunlap Institute for Astronomy \& Astrophysics at the University of Toronto. 
J.J.R.\ acknowledges support from the Canada Research Chairs (CRC) program, the NSERC Discovery Grant program, the Canada Foundation for Innovation (CFI), and the Qu\'{e}bec Minist\`{e}re de l’\'{E}conomie et de l’Innovation.
F.E.B. acknowledges support from ANID-Chile BASAL CATA FB210003, FONDECYT Regular 1241005, and Millennium Science Initiative, AIM23-0001

This work used Jetstream2 cloud computing system at Indiana University through allocation PHY230014 from the Advanced Cyberinfrastructure Coordination Ecosystem: Services \& Support (ACCESS) program, which is supported by U.S. National Science Foundation grants \#2138259, \#2138286, \#2138307, \#2137603, and \#2138296.

Funding for the SDSS and SDSS-II has been provided by the Alfred P. Sloan Foundation, the Participating Institutions, the National Science Foundation, the U.S. Department of Energy, the National Aeronautics and Space Administration, the Japanese Monbukagakusho, the Max Planck Society, and the Higher Education Funding Council for England. The SDSS Web Site is http://www.sdss.org/.

The SDSS is managed by the Astrophysical Research Consortium for the Participating Institutions. 
The Participating Institutions are the American Museum of Natural History, Astrophysical Institute Potsdam, University of Basel, University of Cambridge, Case Western Reserve University, University of Chicago, Drexel University, Fermilab, the Institute for Advanced Study, the Japan Participation Group, Johns Hopkins University, the Joint Institute for Nuclear Astrophysics, the Kavli Institute for Particle Astrophysics and Cosmology, the Korean Scientist Group, the Chinese Academy of Sciences (LAMOST), Los Alamos National Laboratory, the Max-Planck-Institute for Astronomy (MPIA), the Max-Planck-Institute for Astrophysics (MPA), New Mexico State University, Ohio State University, University of Pittsburgh, University of Portsmouth, Princeton University, the United States Naval Observatory, and the University of Washington.

Funding for the Sloan Digital Sky Survey IV has been provided by the Alfred P. Sloan Foundation, the U.S. Department of Energy Office of Science, and the Participating Institutions. 
SDSS-IV acknowledges support and resources from the Center for High Performance Computing  at the University of Utah. The SDSS website is www.sdss.org.

SDSS-IV is managed by the Astrophysical Research Consortium for the Participating Institutions of the SDSS Collaboration including the Brazilian Participation Group, the Carnegie Institution for Science, Carnegie Mellon University, Center for Astrophysics | Harvard \& Smithsonian, the Chilean Participation Group, the French Participation Group, Instituto de Astrof\'isica de Canarias, The Johns Hopkins University, Kavli Institute for the Physics and Mathematics of the Universe (IPMU) / University of Tokyo, the Korean Participation Group, Lawrence Berkeley National Laboratory, Leibniz Institut f\"ur Astrophysik Potsdam (AIP),  Max-Planck-Institut f\"ur Astronomie (MPIA Heidelberg), Max-Planck-Institut f\"ur Astrophysik (MPA Garching), Max-Planck-Institut f\"ur Extraterrestrische Physik (MPE), National Astronomical Observatories of China, New Mexico State University, New York University, University of Notre Dame, Observat\'ario Nacional / MCTI, The Ohio State University, Pennsylvania State University, Shanghai Astronomical Observatory, United Kingdom Participation Group, Universidad Nacional Aut\'onoma de M\'exico, University of Arizona, University of Colorado Boulder, University of Oxford, University of Portsmouth, University of Utah, University of Virginia, University of Washington, University of Wisconsin, Vanderbilt University, and Yale University.

The Pan-STARRS1 Surveys (PS1) and the PS1 public science archive have been made possible through contributions by the Institute for Astronomy, the University of Hawaii, the Pan-STARRS Project Office, the Max-Planck Society and its participating institutes, the Max Planck Institute for Astronomy, Heidelberg and the Max Planck Institute for Extraterrestrial Physics, Garching, The Johns Hopkins University, Durham University, the University of Edinburgh, the Queen's University Belfast, the Harvard-Smithsonian Center for Astrophysics, the Las Cumbres Observatory Global Telescope Network Incorporated, the National Central University of Taiwan, the Space Telescope Science Institute, the National Aeronautics and Space Administration under Grant No. NNX08AR22G issued through the Planetary Science Division of the NASA Science Mission Directorate, the National Science Foundation Grant No. AST-1238877, the University of Maryland, Eotvos Lorand University (ELTE), the Los Alamos National Laboratory, and the Gordon and Betty Moore Foundation.

\software{pandas \citep{pandas2010},
         numpy \citep{numpy2020},    
         scipy \citep{scipy2020},
         matplotlib \citep{matplotlib2007},
         astropy \citep{astropy2013, astropycollaboration2022},  
         emcee \citep{foreman-mackey2013},
         eztao \citep{yu2022}
         }

\bibliography{My_Library}{}
\bibliographystyle{aasjournal}

\appendix
\section{Definitions of DHO-related Quantities}\label{appendix:dho_feats}
Based the classification of a DHO process (underdamped vs. overdamped), the roots of the characteristic equation on the LHS of Equation~\ref{eqn:dho1} (and/or Equation~\ref{eqn:dho2}) can be either real or complex, thus giving different derived timescales and features. 
Here, we list the variety of DHO timescales and features introduced in \citetalias{moreno2019} (updated in \citetalias{yu2022b}).

\noindent The long-term asymptotic amplitude is:
\begin{equation}\label{eqn:dho_amp}
    \sigma_{\mathrm{DHO}} = \sqrt{\frac{\beta_{1}^{2}\alpha_{2} + \beta_{0}^{2}}{2\alpha_{1}\alpha_{2}}}.
\end{equation}

\noindent The roots of the characteristic equation of the LHS of Equation~\ref{eqn:dho1}/Equation~\ref{eqn:dho2} are:
\begin{equation}\label{eqn: dho_char}
    r_{1}, r_{2} = -\frac{\alpha_1}{2} \pm \sqrt{\frac{\alpha_{1}^2}{4} - \alpha_{2}} = -\omega_{0}\xi \pm \omega_{0}\sqrt{\xi^2 - 1}.
\end{equation}

\noindent The characteristic timescales of underdamped DHOs ($\xi < 1$) derived from $r_{1}$ and $r_{2}$ are:
\begin{eqnarray}
    \tau_{\mathrm{decay}} &&= \frac{1}{|\mathrm{Re} (r_{1})|} = \frac{1}{\omega_{0}\,\xi},\\
    T_{\mathrm{dQPO}} &&= \frac{2\pi}{|\mathrm{Im} (r_{1})|} 
    = \frac{2\pi}{\omega_{0}\sqrt{1 - \xi^2}}.
\end{eqnarray}

\noindent The characteristic timescales of overdamped DHOs ($\xi > 1$) derived from $r_{1}$ and $r_{2}$ are:
\begin{eqnarray}
    \tau_{\mathrm{rise}} = |\frac{1}{\mathrm{min}(r_{1}, r_{2})}|,\\
    \tau_{\mathrm{decay}} = |\frac{1}{\mathrm{max}(r_{1}, r_{2})}|.
\end{eqnarray}

\noindent The autocovariance function of a DHO process is:
\begin{equation}\label{eqn:dho_acvf1}
    {\rm ACF}(\Delta t) = A_{1}e^{r_{1}\Delta t} + A_{2}e^{r_{2}\Delta t},
\end{equation}
where
\begin{eqnarray}\label{eqn:dho_acvf2}
    A_{1} = \frac{(\beta_{0}+\beta_{1}r_{1})(\beta_{0}-\beta_{1}r_{1})}{-2\,\mathrm{Re}(r_{1})*(r_{2}-r_{1})(r_{2}^{*}+r_{1})}, \nonumber\\
    A_{2} = \frac{(\beta_{0}+\beta_{1}r_{2})(\beta_{0}-\beta_{1}r_{2})}{-2\,\mathrm{Re}(r_{2})*(r_{1}-r_{2})(r_{1}^{*}+r_{2})}.
\end{eqnarray}

\noindent The structure function is:
\begin{equation}
    {\rm SF}_{\rm DHO}(\Delta t)=\sqrt{2 \sigma_{\text {DHO}}^2(1-{\rm ACF}(\Delta t))}.
\end{equation}\label{eqn:dho_sf}

\end{CJK*}
\end{document}